\documentclass[]{AIAA}
\usepackage{amsmath}
\usepackage{graphicx}
\usepackage{enumerate}
\usepackage{color}
\usepackage{verbatim}
\usepackage[normalem]{ulem}

\begin{document}

\title{Scaling and performance of simultaneously heaving and pitching foils}

\author{Tyler Van Buren\footnote{Research specialist, Mechanical and Aerospace Engineering, Princeton NJ.},
 Daniel Floryan\footnote{Graduate student, Mechanical and Aerospace Engineering, Princeton NJ, AIAA Member.},
 and Alexander J. Smits\footnote{Eugene Higgins Professor, Mechanical and Aerospace Engineering, Princeton NJ, AIAA Fellow.}}
\affiliation{Princeton University, Princeton, New Jersey, 08544}

\begin{abstract}
We consider the propulsive performance of an unsteady heaving and pitching foil, experimentally studying an extensive parameter space of motion amplitudes, frequencies, and phase offsets between the heave and pitch motions. The phase offset $\phi$ between the heaving and pitching motions proves to be a critical parameter in determining the dynamics of the foil and its propulsive performance. To maximize thrust, the heave and pitch motions need to be nearly in phase ($\phi=330^\circ$), but to maximize efficiency, the pitch motion needs to lag the heave motion ($\phi=270^\circ$), corresponding to slicing motions with a minimal angle of attack. We also present scaling relations, developed from lift-based and added mass forces, which collapse our experimental data. Using the scaling relations as a guide, we find increases in performance when foil amplitudes (specifically pitch) increase while maintaining a modest angle of attack. 
\end{abstract}

\maketitle

\section*{Nomenclature}
\noindent\begin{tabular}{@{}lcl@{}}
\textit{$a$}				&=&	trailing edge position \\
\textit{$a^*$}				&=&	non-dimensional trailing edge position, $a^*=a/c$ \\
\textit{$a_h^*$}			&=&	non-dimensional trailing edge position contribution from heave, $a_h^*=h_0/c$ \\
\textit{$a_\theta^*$}		&=&	non-dimensional trailing edge position contribution from pitch, $a_\theta^*=\theta_0$ \\
\textit{$c$}				&=&	foil chord \\
\textit{$C_L$}				&=&	lift coefficient, $C_L$   \\
\textit{$C_T$}				&=&	thrust coefficient, $C_T=F_x/\frac{1}{2}\rho U_\infty^2sc$   \\
\textit{$C_P$}				&=& power coefficient, $C_P=P/\frac{1}{2}\rho U_\infty^3sc$ \\
\textit{$f$}				&=&	frequency of motion \\
\textit{$f^*$}				&=&	reduced frequency, $f^*=fc/U_\infty$\\
\textit{$F_D$}				&=&	drag offset \\
\textit{$F_{x,y}$}			&=&	streamwise ($x$) and lateral ($y$) force \\
\textit{$h$}				&=& heave position (leading edge) \\
\textit{$L$}				&=& Lift force\\
\textit{$m_w$}				&=&	mass of displaced fluid in the wake\\
\textit{$M_z$}				&=&	spanwise moment \\
\textit{$P$}				&=&	foil input power, $P=F_y\dot{h} + M_z\dot{\theta}$  \\
\textit{$Re$}				&=&	chord based Reynolds number, $Re=cU_\infty/\nu$ \\
\textit{$s$}				&=&	foil span \\
\textit{$St$}				&=&	Strouhal number, $St=2fa_0/U_\infty$\\
\textit{$St_h$}				&=&	Strouhal number based on heave, $St=2fh_0/U_\infty$\\
\textit{$St_\theta$}		&=&	Strouhal number based on pitch, $St=2fc\theta_0/U_\infty$\\
\textit{$t$}				&=&	time \\
\textit{$u_w$}				&=&	characteristic wake velocity \\
\textit{$U_\infty$}			&=&	freestream velocity \\
\textit{$U_{\text{eff}}$}	&=&	effective velocity incoming to foil, $U_{\text{eff}}=\sqrt{U_\infty^2+\dot h^2}$ \\
\textit{$U^*$}				&=&	non-dimensional effective velocity, $U^*=\overline{U_{\text{eff}}}/U_\infty$ \\
\end{tabular}

\noindent\begin{tabular}{@{}lcl@{}}
\textit{$\alpha$}			&=&	angle of attack \\
\textit{$\eta$}				&=& propulsive efficiency, $\eta=C_T/C_P$\\
\textit{$\theta$}			&=&	pitch angle \\
\textit{$\nu$}				&=&	kinematic viscosity \\
\textit{$\rho$}				&=&	fluid density \\
\textit{$\phi$}				&=&	phase angle between heave and pitch motions \\
\textit{$\psi$}				&=&	phase angle between heave and angle of attack motions \\
\textit{$_0$}				&=& (subscript) amplitude of time varying signal \\
\textit{$\dot{ }$}			&=& (accent) first time derivative \\
\textit{$\ddot{ }$}			&=& (accent) second time derivative \\
\textit{$\overline{\phantom{A}}$}		&=& (accent) time average \\
\end{tabular} \\

\section{Introduction}
In the past several decades there has been a concerted effort to find new methods of propulsion for underwater vehicles that are inspired by aquatic life \cite{sfakiotakis1999, triantafyllou2000, wu2011}.  Many fish swim using oscillatory propulsion methods \cite{lindsey1978}, where the principal thrust comes from oscillating a propulsive surface, such as a fluke or caudal fin, in a combined heave and pitch motion.   In many cases a clear distinction can be made between the body as the main source of drag, and propulsor as the main source of thrust, and then the two may be effectively studied in isolation.  
Here, we focus our attention on the thrust and efficiency of a foil moving in heave and pitch as a simplified model of an isolated propulsor,  for possible application to a new generation of underwater vehicles.  

The performance of submerged foils in combined heaving and pitching motion has already been studied relatively extensively \cite{lighthill1970,dickinson1996unsteady,sfakiotakis1999,triantafyllou2000,von2003flow}. In a particularly influential work, Anderson et al. \cite{anderson1998} obtained efficiencies as high as 87\% using a heaving and pitching two-dimensional NACA 0012 airfoil in sinusoidal motion.  They connected the performance to the wake produced by the foil, arguing that for maximum efficiency the leading edge vortex pair needs to beneficially interact with the trailing-edge vorticity.  In related work, Read et al. \cite{read2003} recognized the importance of the peak angle of attack when considering performance, though they did find conflicting values of efficiency for similar foil motions (55-70\%) when compared to \cite{anderson1998}. The phase angle between the heave and pitch motions was also examined briefly, although the authors, somewhat surprisingly, did not find it to be very influential. Experiments on large amplitude motions by Scherer \cite{scherer1968} showed similar peak efficiency values to \cite{read2003} for a wide range of parameters.

In terms of optimization, Kaya and Tuncer \cite{kaya2007} numerically studied heaving and pitching airfoil performance in laminar air flow, and used gradient-based optimization of the motion paths to show that there were significant benefits to thrust by moving non-sinusoidally.  They found that motions that maintain a constant angle of attack for longer periods of time, a topic also considered by \cite{read2003}, produced higher thrust than purely sinusoidal motions. 

The theory of heaving and pitching plates and foils has a long history. Theodorsen \cite{theodorsen1935} derived the linearized theory for forces on an oscillating foil in the context of aerodynamic flutter. This derivation included linearized added mass, though using Sedov \cite{sedov1965} one could include the nonlinear contribution (as we do in this study). Garrick \cite{garrick1937} extended \cite{theodorsen1935} to the context of propulsion by introducing expressions for thrust and power for a two-dimensional, rigid propulsor. Lighthill \cite{lighthill1970} used \cite{garrick1937} to estimate the forces produced by the lunate tail of a fish in the context of his \emph{elongated-body theory}, which was later extended to large amplitude motions by Lighthill \cite{lighthill1971} (although only added mass forces were considered in the latter work). Chopra \cite{chopra1974} and Chopra and Kambe \cite{chopra1977} made \cite{lighthill1970} three-dimensional by incorporating lifting-line theory. Katz and Weihs \cite{katz1978} was the first to consider the addition of chordwise flexibility on a two-dimensional foil where the solid mechanics were coupled to the fluid mechanics (Wu \cite{wu1961} earlier considered a flexible foil, though the deformation of the foil was imposed rather than governed by the fluid and elastic solid forces).

What appears to be missing in the literature is an approachable scaling analysis that takes into account the possibly nonlinear underlying physical mechanisms that generate thrust and determine efficiency. In this respect, Floryan et al. \cite{Floryan2016} combined unsteady lift \cite{theodorsen1935} and added mass forces \cite{sedov1965} to construct scaling relations which describe the mean forces generated by heaving \emph{or} pitching foils. They showed that the mean thrust generated by heaving motions is entirely lift-based, whereas mean thrust generated by pitching motions is from added mass alone. Conversely, for both heave and pitch, the mean input power (and thus efficiency) depends on both lift-based and added mass forces. These scaling relations have since been extended to intermittent motions \cite{Floryan2017} by accounting for the duty cycle of the motion, and to non-sinusoidal motions \cite{VanBuren2017} by adding a parameter based on the peak trailing edge velocity of the foil. 

Here, we extend the approach of \cite{Floryan2016} to derive scaling relationships for foils that are simultaneously heaving and pitching.  We verify the relationships against  experiments covering a large parameter space in heave and pitch amplitudes, frequencies, and phase differences.  This approach allows us to pinpoint physical mechanisms that influence thrust and efficiency, which we can use as a guide to achieve higher performance. 

\section{Experimental methods}
\begin{figure}
\centering
\includegraphics[width=0.75\textwidth]{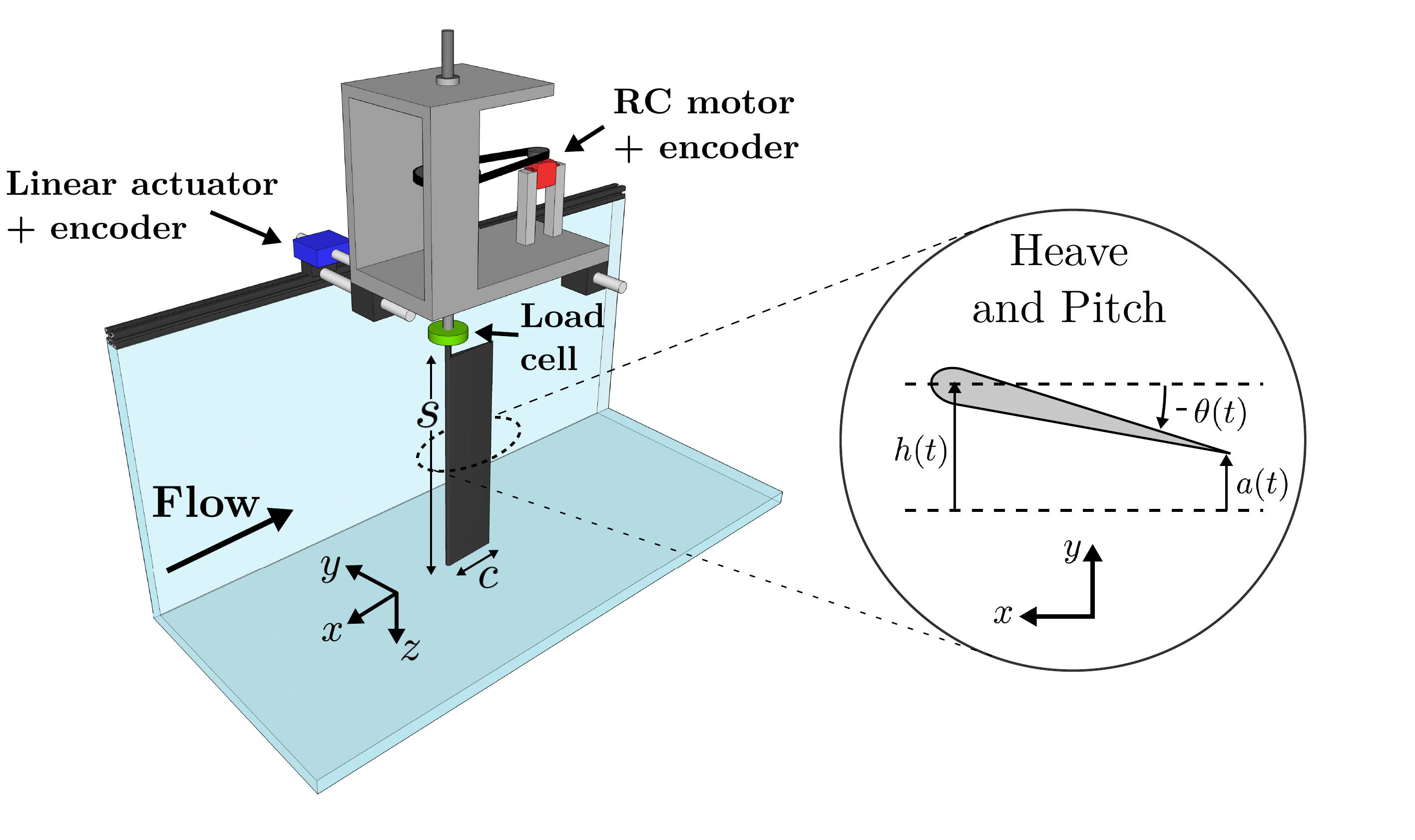}
\caption{Experimental setup and motion definition.}
\label{fig:expSetup} 
\end{figure}

Experiments on a simultaneously heaving and pitching foil were conducted in a free-surface recirculating water tunnel with a 0.46 m $\times$ 0.3 m $\times$ 2.44 m test section and baffles to minimize surface waves. The tunnel velocity was fixed at a constant value $U_\infty=0.1$ m/s. The experimental setup is illustrated in figure \ref{fig:expSetup}.

\begin{figure}
\centering
\includegraphics[width=0.5\textwidth]{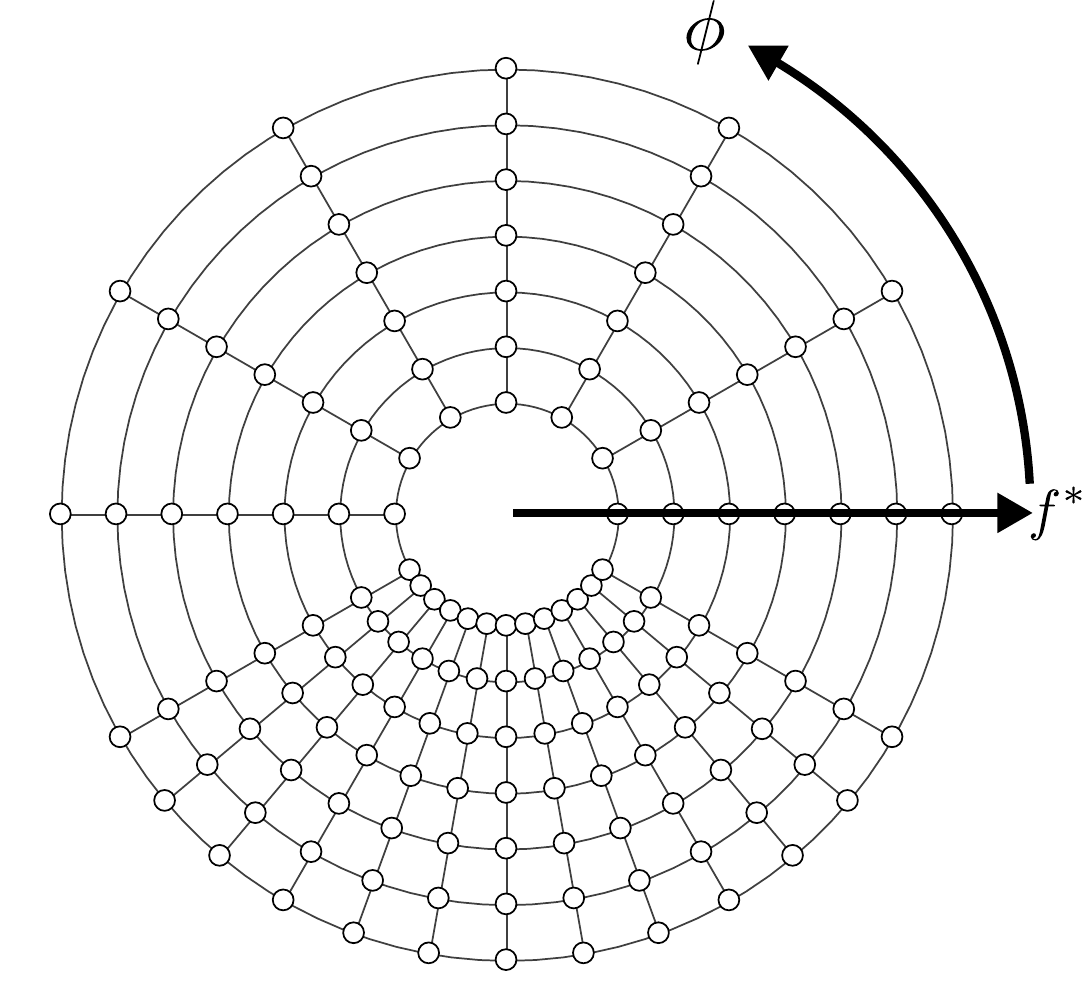}
\caption{Diagram depicting the parameter space on a frequency-phase plot. }
\label{fig:phaseCases}
\end{figure}

A teardrop foil was used with a chord $c=80$ mm, maximum thickness of 8 mm, and span $s=279$ mm, yielding a chord-based Reynolds number of $Re=8000$. Heave motions were generated by a linear actuator (Linmot PS01-23x80F-HP-R) pushing the foil carriage on near-frictionless air bearings (NewWay S301901), and pitch motions about the leading edge were generated by a servo motor (Hitec HS-8370TH).  
Both motions were simultaneously measured via encoders. The foil was actuated sinusoidally according to 
\begin{equation}
h = h_0 \sin(2 \pi f t),\quad
\theta = \theta_0 \sin(2 \pi f t+ \phi).
\label{eq:motions}
\end{equation} 
at frequencies $f=0.2$ Hz to 0.8 Hz every 0.1 Hz, with heave amplitudes $h_0=10$ mm, 20 mm, and 30 mm, pitch amplitudes $\theta_0=5^\circ$, $10^\circ$, and $15^\circ$, and phase differences $\phi=0^\circ$ to 330$^\circ$ in intervals of 30$^\circ$, with a more refined spacing of 10$^\circ$ between 210$^\circ$ and 330$^\circ$ (see table \ref{tab:Cases} and figure \ref{fig:phaseCases}). 
Altogether, the parameter space comprised 1260 unique cases. 

\begin{table}
\begin{center}
\begin{tabular}{l l}
 Parameter & Range\\ \hline
 Freestream velocity & $U_\infty=0.1$ m/s \\
 Chord & $c=80$ mm \\
 Span & $s=279$ mm \\
 Frequency & $f=0.2, 0.3, \ldots, 0.8$ Hz \\
 Heave amplitude   & $h_0=10, 20, 30$ mm \\
  Pitch amplitude   & $\theta_0=5^\circ, 10^\circ, 15^\circ$ \\
 Phase offset	    & $\phi = 0^\circ, 30^\circ, \ldots, 210^\circ, 220^\circ, \ldots, 330^\circ$ \\
\end{tabular}
\caption{Experimental parameter space.}
\label{tab:Cases}
\end{center}
\end{table}

The forces and moments of the foil were measured using a six-component sensor (ATI Mini40), with force and torque resolutions of $5\times 10^{-3}$ N and $1.25\times 10^{-4}$ N m, respectively, in the $x$- and $y$-directions and $10^{-2}$ N and $1.25\times 10^{-4}$ N m in the $z$-direction, sampled at 100 Hz. Each case ran for 30 cycles of the motion, with the first and last 5 cycles used for warm-up and cool-down. Before every case we zeroed the force sensor to minimize any voltage drift. Due to the sufficient repeatability shown by similar experiments in the past \cite{Floryan2016, Floryan2017, VanBuren2017}, only one trial of each case was performed. 

\section{Results and discussion}
We present the results on propulsive performance in terms of the non-dimensional thrust coefficient, input power coefficient, and Froude efficiency, defined by
\begin{equation}
C_T = \frac{F_x}{\frac{1}{2}\rho U_\infty^2sc},\qquad C_P=\frac{F_y\dot{h}+M_z\dot{\theta}}{\frac{1}{2}\rho U_\infty^3sc},\qquad \eta=\frac{C_T}{C_P}.
\label{perfEQ}
\end{equation}
We denote the time-averaged values of these parameters by $\overline{C}_T$ and $\overline{C}_P$, respectively, while efficiency is always reported as a time-averaged quantity. The foil kinematics are characterized by the Strouhal, $St = 2fa_0/U_\infty$, where $a_0$ is the peak amplitude of the trailing edge motion, and by the reduced frequency, $f^* = fc/U_\infty$. 

\begin{figure}
\centering
\includegraphics[width=1\textwidth]{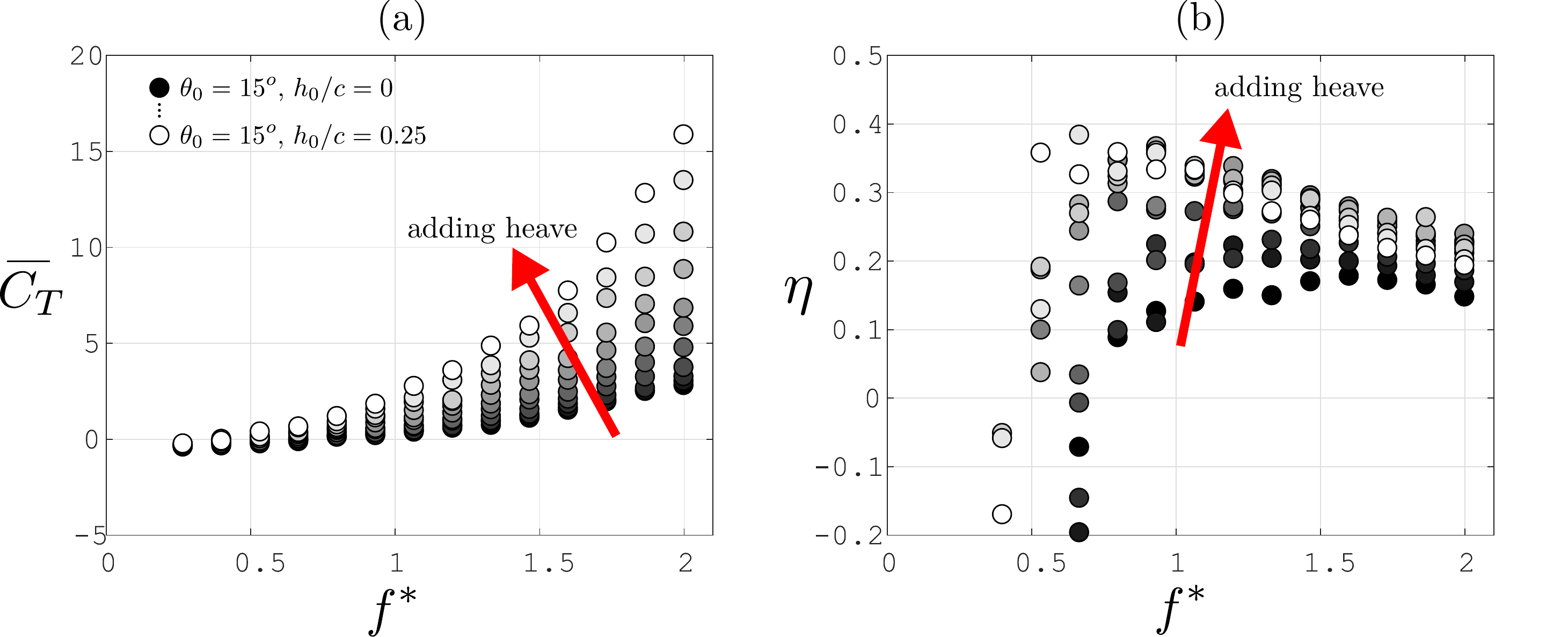}
\caption{Time-averaged (a) thrust coefficient and (b) efficiency of a pitching foil with incremental increases in heave amplitude.  Phase offset between heave and pitch is $\phi=270^\circ$. Experimental conditions specific to this figure: $U_\infty=60$ mm/s, $f=0.2,0.3, \ldots, 1.5$ Hz, $h_0=0,2, \ldots, 20$ mm.}
\label{fig:pitchHeave}
\end{figure}

To illustrate the effects of combining heave and pitch motions, figure~\ref{fig:pitchHeave} shows the time-averaged thrust coefficient and efficiency of a pitching foil with incremental increases in heave amplitude while keeping the pitch amplitude fixed at $\theta_0=15^\circ$ (the experimental results in figures~\ref{fig:pitchHeave}a and ~\ref{fig:pitchHeave}b are from a different dataset and differ from table \ref{tab:Cases}). In this case, adding heave to the pitching motion increases thrust and efficiency significantly.  Note also that the efficiency curves exhibit a maximum value, indicating that there is an optimum efficiency point. For low values of $f^*$, the efficiency decreases sharply as the effects of the viscous drag on the propulsor become important. For high values of $f^*$, the efficiency approaches its inviscid or ideal value, which slowly decreases with increasing reduced frequency.

It should be noted that we are not the first to suggest the necessity of combining pitching and heaving motions, which is first discussed in von K{\'a}rm{\'a}n and Burgers \cite{vonKarman1934} and more recently recapitulated by Wu \cite{wu2011}. Our results reinforce this observation.

\begin{figure}
\centering
\includegraphics[width=0.5\textwidth]{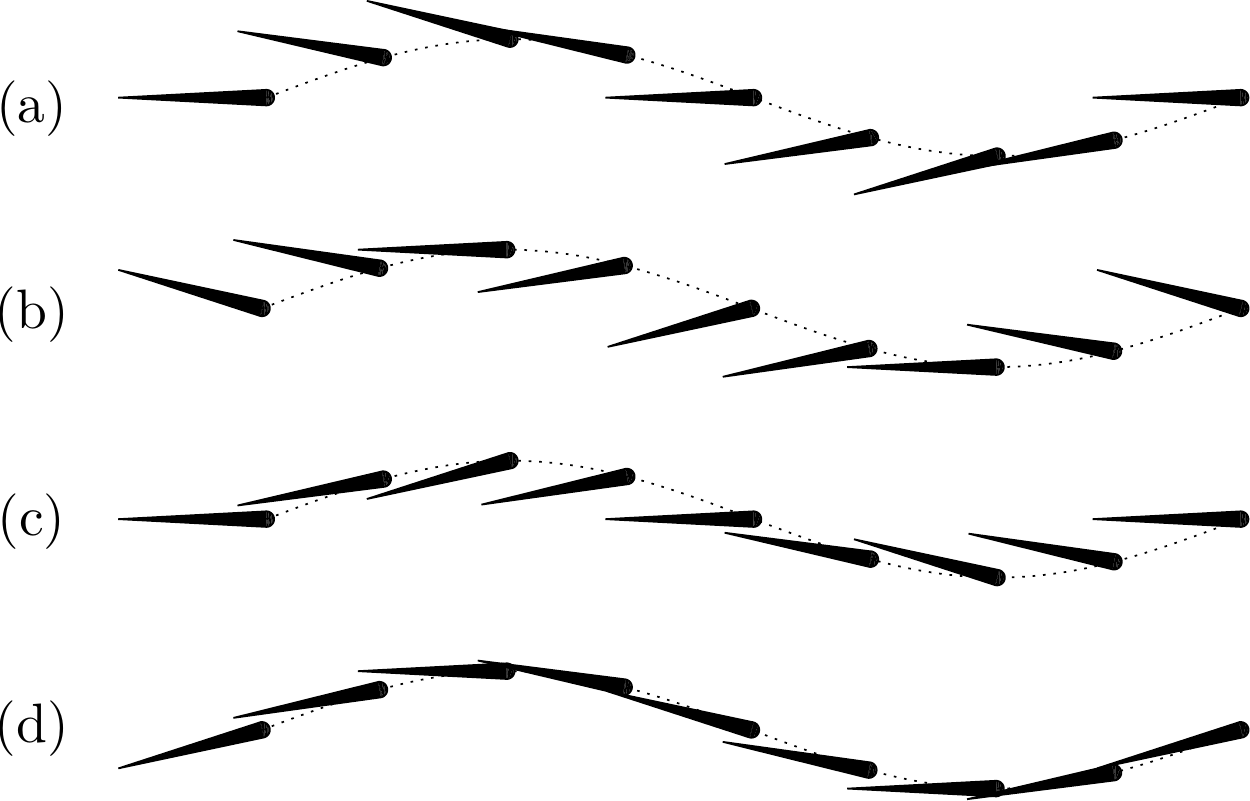}
\caption{Motion of a foil swimming from left to right via heave and pitch motions with a phase offset (a) $\phi=0^\circ$, (b) $90^\circ$, (c) $180^\circ$, and (d) $270^\circ$. In this example, $h_0/c=0.375$, $\theta_0=15^\circ$, and $f^*=0.16$.}
\label{fig:foilMotion}
\end{figure}

\subsection{Maximizing performance through combined motion}
When combining sinusoidal heaving and pitching motions, the phase offset becomes a critical parameter. Figure \ref{fig:foilMotion} illustrates the motion of a swimming foil for phase differences $\phi=0^\circ, 90^\circ, 180^\circ,$ and $270^\circ$. When heave and pitch are in phase, the motion to an observer is as if the foil is pitching about some point upstream of the leading edge. For phase angles around  $\phi=90^\circ$, the trailing edge leads the leading edge. When $\phi=180^\circ$ the foil appears to pitch about a point behind the leading edge. However, $\phi=270^\circ$ seems to be the most ``fish-like'', cleanly slicing through the water with the lowest angles of attack (represented by the angle between the foil and the local tangent of the dotted line).

\begin{figure}
\centering
\includegraphics[width=0.9\textwidth]{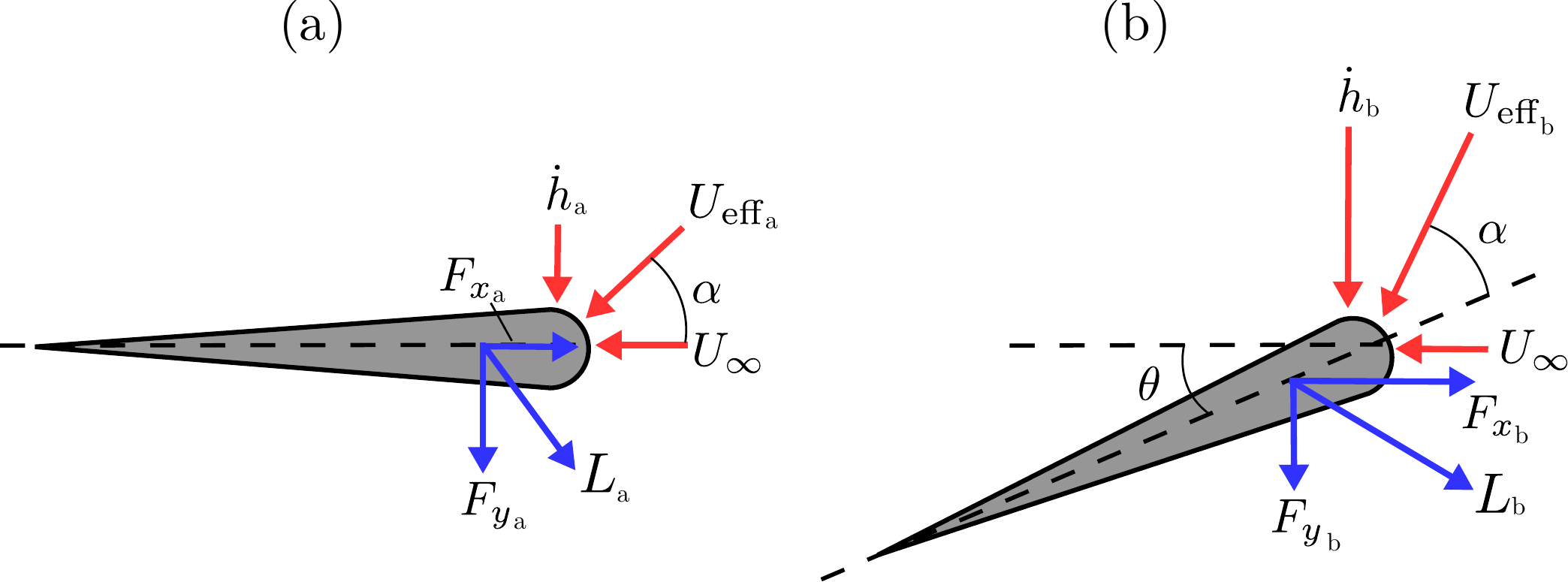}
\caption{Heaving foil (a) without  and (b) with added pitch motion. Streamwise, heave, and effective velocities shown in red, resulting lift-based forces shown in blue.}
\label{fig:bigTheta}
\end{figure}

To appreciate the importance of pitch angle in regards to thrust production, we must first consider how thrust is generated from heave-only motions, as shown in the diagram in figure \ref{fig:bigTheta}a. Thrust production in heave is lift-based, meaning that the lateral velocity of the foil $\dot{h}$ combines with the free-stream velocity $U_\infty$ into an effective foil velocity $U_\text{eff}$ that produces lift $L$. Since the lift vector is perpendicular to the effective velocity vector, portions of $L$ are in the lateral and streamwise directions, $F_y$ and $F_x$ respectively, with the latter being the thrust. As in steady aerodynamics, to efficiently produce thrust we want to avoid dramatic separation, thus we want modest angles of attack $\alpha$. This is a performance limitation, because in order to produce more thrust we need to increase the heave velocity, which in turn increases the angle of attack.

This problem can be mitigated by adding pitch such that it counters the angle of attack induced by heave. Consider a heaving foil with ``fish-like'' pitching motion added ($\phi=270^\circ$) as shown in the diagram in figure \ref{fig:bigTheta}b. Comparing to figure \ref{fig:bigTheta}a, the heave velocity is greater but the angle of attack is the same due to the addition of appropriate pitch. The increased heave velocity both increases the lift and rotates more of the lift vector in the thrust direction, more efficiently producing thrust. Thus, we see the importance of combining the two motions and having an appropriate phase between them.

\begin{figure}
\centering
\includegraphics[width=0.8\textwidth]{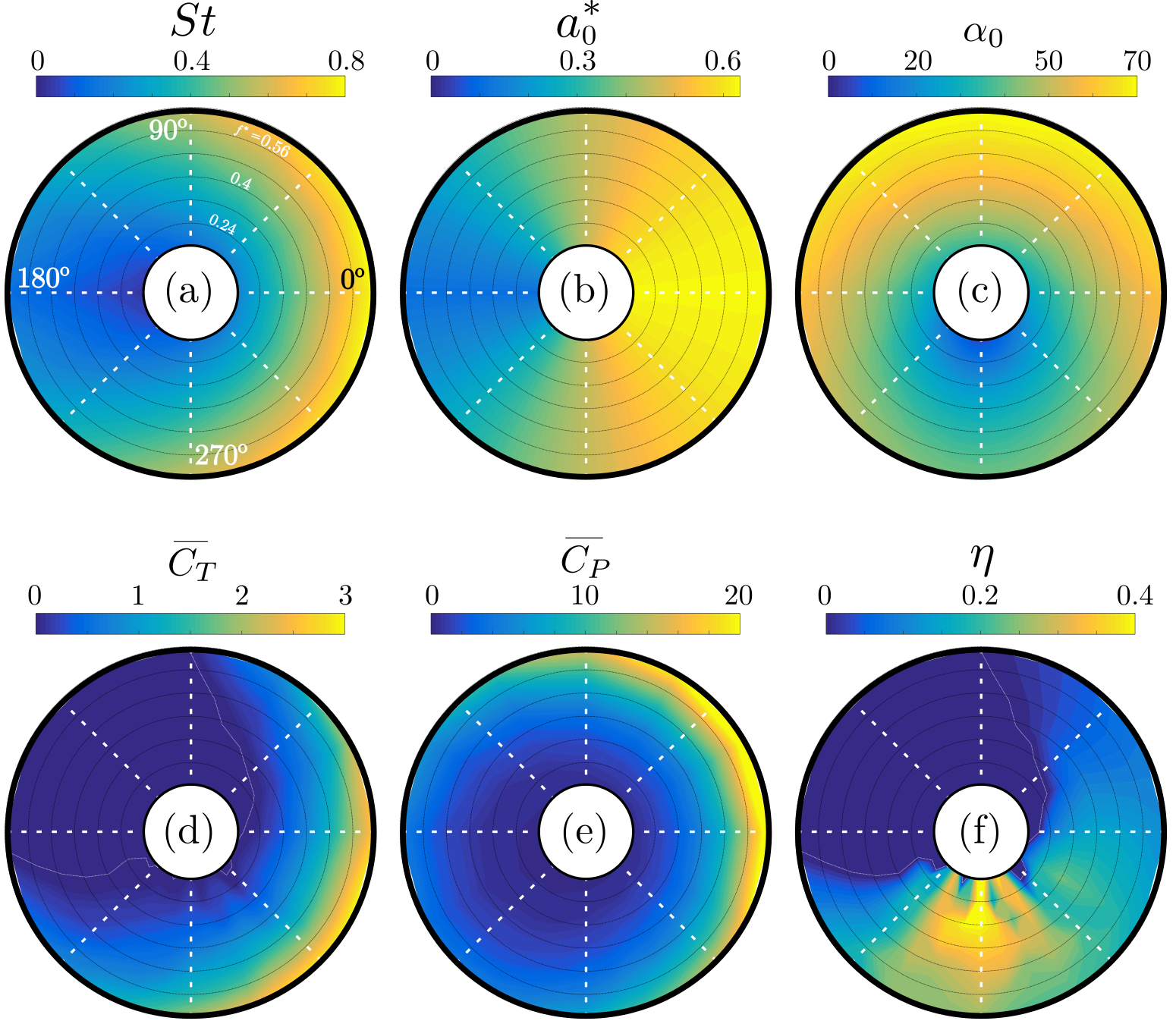}
\caption{Impact of phase offset ($\phi$, shown in the azimuthal variation) and reduced frequency ($f^*$, shown in the radial direction) on: (a) Strouhal number; (b) amplitude to chord ratio; (c) peak angle of attack; (d) time-averaged thrust coefficient; (e) time-averaged power coefficient; and (f) efficiency. Heave amplitude to chord ratio $h_0/c=0.375$ and pitch amplitude $\theta_0=15^\circ$. Reduced frequency increases radially outward with lines marking levels at $f^*=0.16, 0.24, \ldots, 0.64$. }
\label{fig:phases}
\end{figure}

Figure \ref{fig:phases} shows the time-averaged output performance (thrust, power, and efficiency) for all the phases and frequencies tested at heave and pitch amplitudes of $h_0/c=0.375$ and $\theta_0=15^\circ$, respectively. In each subfigure, the phase difference $\phi$ varies in the azimuthal direction, while the reduced frequency $f^*$ varies in the radial direction.  Note that these trends are representative of all of the combinations of heave and pitch amplitudes tested in this study, and that the full results are presented in Appendix A.  

The Strouhal number $St$ and trailing edge amplitude $a_0^*$ are largest for $\phi=0^\circ$, since in-phase heave and pitch motions will result in the greatest peak-to-peak excursion of the trailing edge. The peak angle of attack $\alpha_0$, however, aligns with $\phi=90^\circ$. Interestingly, the thrust does not match the behavior of the Strouhal number, which would be expected if the Strouhal number is the governing non-dimensional parameter of these flows, as suggested by \cite{triantafyllou1993}.  The peak thrust actually occurs around $\phi\approx330^\circ$, where the trailing edge of the foil lags the leading edge by $30^\circ$.  Conversely, the peak power is tilted towards $\phi\approx30^\circ$. The peak efficiency occurs around $\phi=270^\circ$, coincident with the smallest peak angles of attack. This observation agrees with other researchers who have argued that the peak angle of attack is an important performance parameter \cite{anderson1998, kaya2007}, with the former study specifically using $\phi=270^\circ$ to achieve the highest efficiencies. We will explore this further in section \ref{S3S2}.

\begin{figure}
\centering
\includegraphics[width=0.7\textwidth]{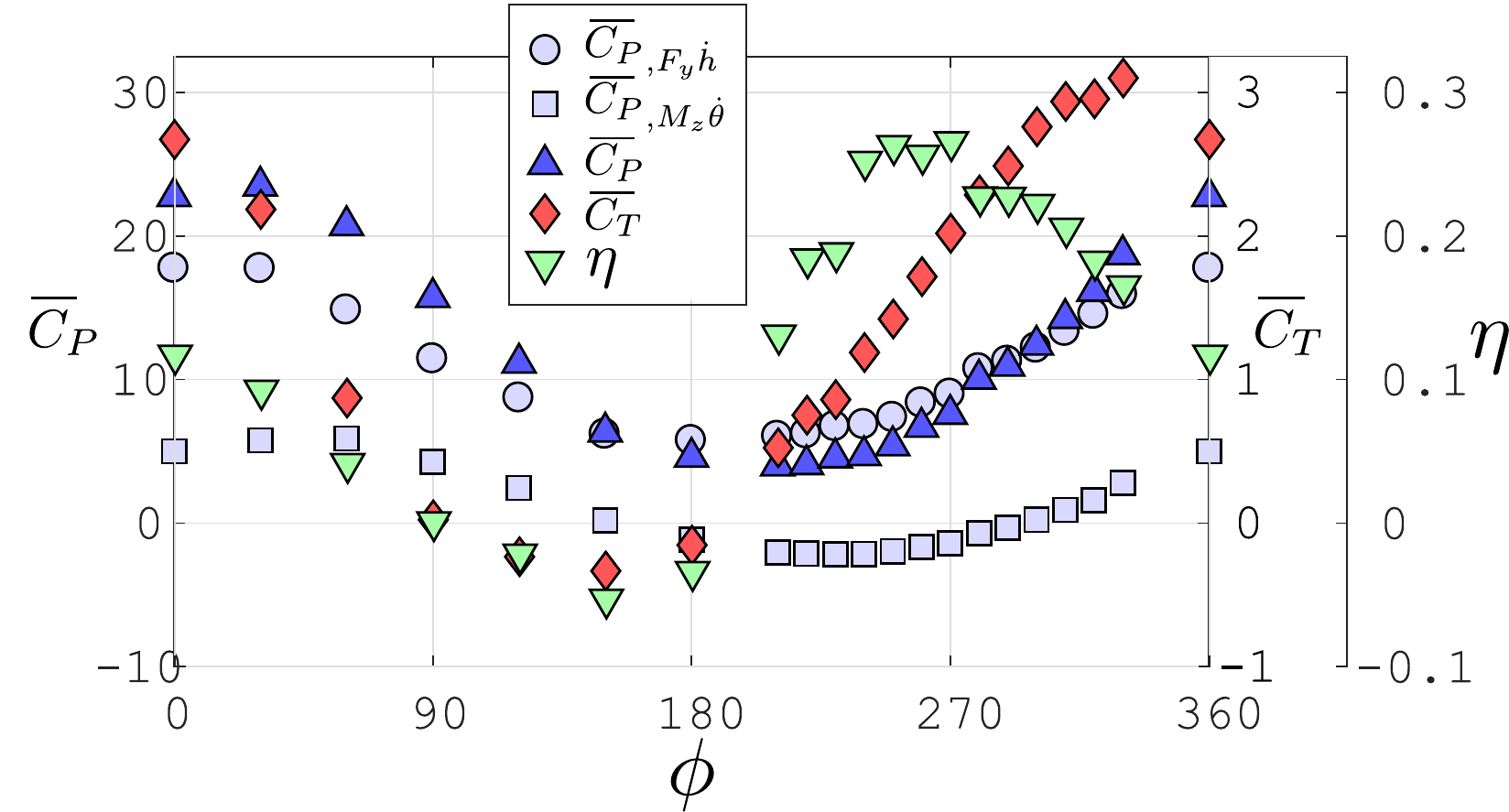}
\caption{Time-averaged power coefficients (force component, moment component, and total), thrust coefficient, and efficiency as they vary with phase between heave and pitch motions. Heave amplitude to chord ratio $h_0/c=0.375$, pitch amplitude $\theta_0=15^\circ$, and reduced frequency $f^*=0.64$.}
\label{fig:powerComponent}
\end{figure}

The impact of phase offset can be more clearly seen in figure \ref{fig:powerComponent}, which shows the same data displayed in figure \ref{fig:phases}, but for a fixed reduced frequency of $f^*=0.64$. The peak thrust and minimum power occur when $\phi=330^\circ$ and 210$^\circ$, respectively, and the peak efficiency is almost exactly in between at $\phi=270^\circ$. Interestingly, we do not see positive thrust for $90^\circ<\phi<180^\circ$, even at the highest frequency and largest motion amplitudes, indicating this range of phase offsets is propulsively of no value. 

\begin{figure}
\centering
\includegraphics[width=0.7\textwidth]{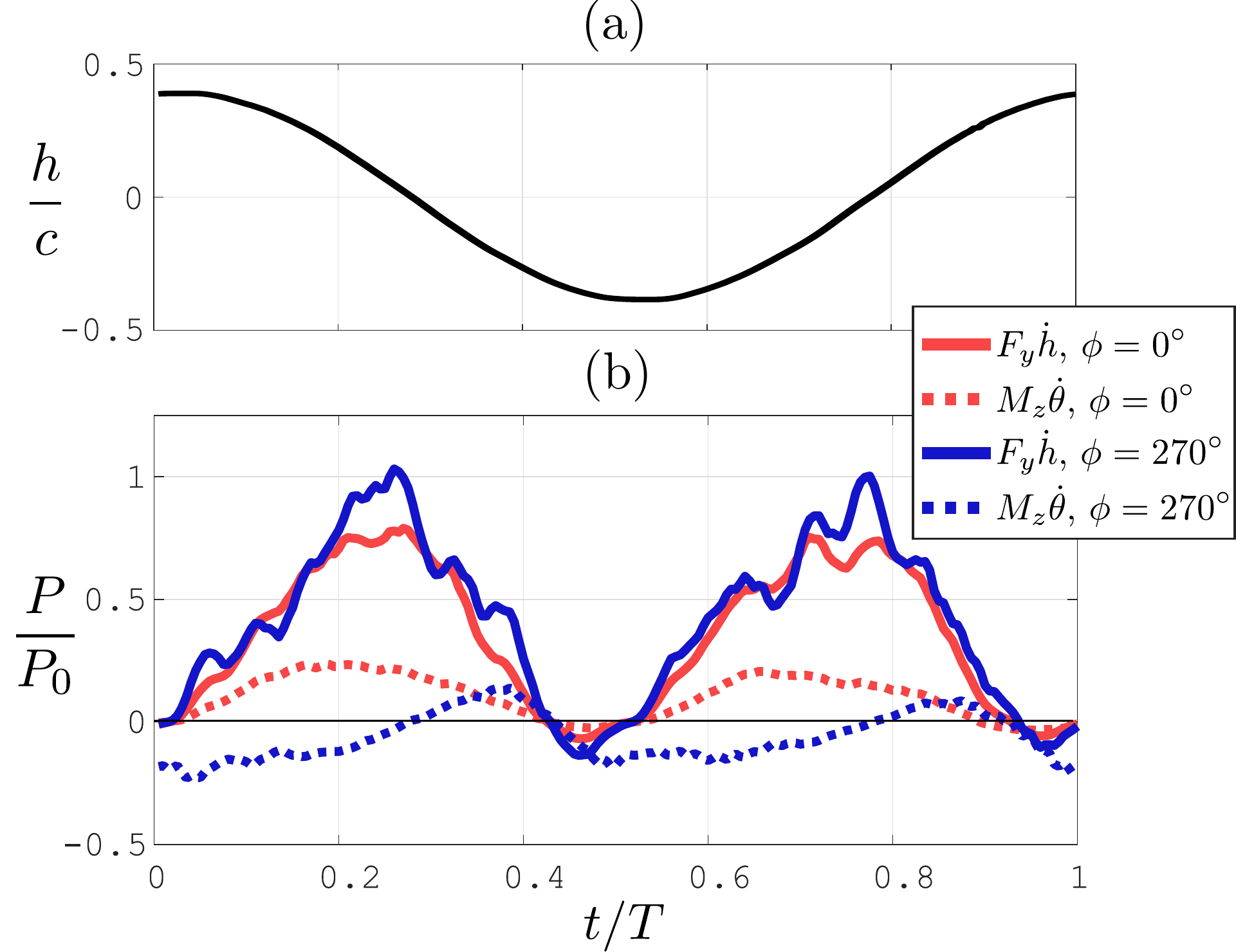}
\caption{Phase-averaged cycles of (a) heave to chord ratio and the (b) force ($F_y \dot h$) and moment ($M_z \dot \theta$) components of power for $\phi=0^\circ$ (red) and 270$^\circ$ (blue).}
\label{fig:timePower}
\end{figure}

To better understand the input power behavior shown in figure \ref{fig:powerComponent}, we split the power into its lateral force ($F_y \dot h$) and moment ($M_z \dot \theta$) components.  Figure \ref{fig:timePower} shows the relative contributions of the lateral force and moment to the power over one actuation cycle for phase offsets $\phi=0^\circ$ and 270$^\circ$. 
The contributions of the lateral force are similar for both cases, but the moment components differ markedly.  When the heave and pitch motions are in phase ($\phi=0^\circ$), their accelerations are in phase. Thus, when the foil starts to accelerate laterally, it also accelerates rotationally in the same direction, working against the resistance of the fluid, yielding a high moment component in the power. However, when heave and pitch are offset by $\phi=270^\circ$, the lateral acceleration of the foil produces a force that assists the force associated with pitch rotation, thus lowering the moment required. This interaction between the lateral force and moment components of the input power is a critical aspect to achieving the high efficiency we see at $\phi=270^\circ$.

In summary, we find that the most efficient motions are when $\phi=270^\circ$, where the trailing edge lags the leading edge in a slicing motion. In contrast, the highest thrust motions occur when the heaving and pitching motions are nearly in phase, though this comes with a significant loss in efficiency. Our experimental results differ from those calculated using the linear theory \cite{garrick1937}. According to the linear theory, thrust is maximized when heave and pitch are in phase, and efficiency is maximized for phases around $225^\circ$. The discrepancy between our experiments and the linear theory motivates us to develop scaling laws which can explain our observations.

\subsection{Scaling laws}\label{S3S2}

To further understand the forces governing these motions and how we can use this understanding to improve performance, we develop a scaling analysis by considering the forces acting on the foil due to added mass and lift-based sources.  We do not include the forces from the inertia of the foil, as was done in \cite{scherer1968}, because their time-averaged contributions will be exactly zero for these types of flows, as shown in \cite{VanBuren2018}. We base our analysis on the work by Floryan et al. \cite{Floryan2016}, who considered heaving and pitching motions independently. 

Throughout the analysis we make a small angle approximation on the pitch angle. We only consider pitching about the leading edge, however as pointed out by Lighthill \cite{lighthill1970}, changing the pitch axis is equivalent to changing the phase angle between heaving and pitching motion (for small motions), which is considered here. At the end of the section, we will present expressions for thrust and power that will have terms with empirically determined coefficients applied. Thus, throughout the analysis we will be loose with our notation, dropping any constants since they will eventually be embedded in the empirical coefficients. Also note that we use the $\sim$ symbol to indicate that one quantity scales as another (e.g., $F=ma$ for constant mass is $F \sim a$), and we use the $\approx$ symbol to indicate one quantity is approximately equal to another (e.g., 1.013$\approx$1).

First, we consider the lift-based (circulatory) forces as in Theodorsen \cite{theodorsen1935}. Traditional quasi-steady lift force scales as $\sim \rho s c \,U_{\text{eff}}^2\, C_L$ where the lift coefficient scales as $C_L \sim \alpha + \dot{\alpha}c/U_{\text{eff}}$. Projecting the lift force onto the streamwise and lateral directions (multiplying by $\dot{h}/U_{\text{eff}}$ and $U_\infty/U_{\text{eff}}$, respectively) gives
\begin{equation*}
F_{x,L} \sim \rho s c \Big(
\underbrace{\alpha \dot{h}\, U_{\text{eff}} \rule[-12pt]{0pt}{5pt}}_{\mbox{1}} +
\underbrace{c\, \dot{\alpha}\dot{h} \rule[-12pt]{0pt}{5pt}}_{\mbox{2}}
\Big),
\label{eq:liftFx}
\end{equation*}
\begin{equation*}
F_{y,L} \sim \rho s c \Big(
\underbrace{\alpha \, U_\infty U_{\text{eff}} \rule[-12pt]{0pt}{5pt}}_{\mbox{1}} +
\underbrace{c\, \dot{\alpha} \, U_\infty  \rule[-12pt]{0pt}{5pt}}_{\mbox{2}}
\Big),
\label{eq:liftFy}
\end{equation*}
where groups 1 and 2 are the steady and unsteady portions of the lift, respectively. The resulting moment about the leading edge is
\begin{equation*}
M_{z,L} \sim \rho s c^2 U_{\text{eff}} \, \left(
\alpha \, U_{\text{eff}} +
c\, \dot{\alpha} 
\right).
\label{eq:liftMz}
\end{equation*}
The angle of attack is exactly $\alpha=-\theta-\arctan(\dot{h}/U_\infty)$, however, for heave velocities on the order of the freestream velocity or less we can simplify it to $\alpha\approx-\theta-\dot{h}/U_\infty$. At this point we choose to leave lift force terms as functions of $\alpha$ and not $\{\theta,\, h\}$ to easily keep track of lift-based phenomena.

Next we consider the streamwise and lateral forces produced by the added mass forces, following the analysis of Sedov \cite{sedov1965} and transforming into the laboratory frame of reference (same process as in \cite{Floryan2016}). This yields
\begin{equation*}
F_{x,AM} \sim \rho s c^2 \Big(
\underbrace{c \ddot{\theta}\theta +
\ddot{h}\theta \rule[-12pt]{0pt}{5pt}}_{\mbox{1}} +
\underbrace{\dot{h}\dot{\theta} (1+\theta^2)  +
\dot{\theta}\theta\,U_\infty \rule[-12pt]{0pt}{5pt}}_{\mbox{2}} + 
\underbrace{ c\, \dot{\theta}^2 \rule[-12pt]{0pt}{5pt}}_{\mbox{3}}
\Big),
\label{eq:addedmassFx}
\end{equation*}
\begin{equation*}
F_{y,AM} \sim \rho s c^2 \Big(
\underbrace{c\,\ddot{\theta}  +
\ddot{h} \rule[-12pt]{0pt}{5pt}}_{\mbox{1}} +
\underbrace{\dot{h}\dot{\theta}\theta + 
\dot{\theta} (1+\theta^2) \,U_\infty \rule[-12pt]{0pt}{5pt}}_{\mbox{2}} + 
\underbrace{ c\, \dot{\theta}^2 \theta \rule[-12pt]{0pt}{5pt}}_{\mbox{3}} 
\Big),
\label{eq:addedmassFy}
\end{equation*}
where group 1 terms are forces arising from the foil accelerations, group 2 terms are Coriolis forces, and group 3 terms are the centrifugal forces. The moment about the leading edge due to added mass is
\begin{equation*}
M_{z,AM} \sim \rho s c^2 \Big(
c^2 \, \ddot{\theta}  +
c \, \ddot{h}  +
\dot{h} (1+\theta^2) \, U_\infty + 
\theta \, U_\infty^2 + 
\dot{h}^2\theta 
\Big).
\label{eq:addedmassMz}
\end{equation*}
Note that $1+\theta^2$ is of order $\mathcal{O}(1+\theta_0^2)$, so we will make the approximation that $1+\theta^2 \approx 1$.

Combining the contributions from added mass and lift yields the following expression for total thrust
\begin{equation}
F_x \sim \rho s c \left(
\alpha \dot{h}\, U_{\text{eff}} +
c\, \dot{\alpha}\dot{h} + 
c^2 \ddot{\theta}\theta +
c\,\ddot{h}\theta +
c\,\dot{h}\dot{\theta} + 
\underline{c\, \dot{\theta} \theta \, U_\infty} +
c^2 \dot{\theta}^2
\right)-F_D,
\label{eq:thrust}
\end{equation}
where $F_D$ is the fluid drag force on the foil. Similarly, combining the contributions from added mass and lift for the power $P=F_y\dot{h} + M_z \dot{\theta}$ yields
\begin{equation}
\begin{aligned}
P \sim &\rho s c\, \Big(   
\alpha \dot{h} \, U_{\text{eff}} \, U_\infty  + 
c \, \dot{\alpha}\dot{h} \,  U_\infty +
c \, \alpha \dot{\theta} \, U_{\text{eff}}^2 +
c^2 \, \dot{\alpha} \dot{\theta} \, U_{\text{eff}} +
c^2 \dot{h} \ddot{\theta} +
\underline{c\, \ddot{h}\dot{h}} 
\\
& + c\, \dot{h}^2\dot{\theta}\theta +
c \, \dot{h}\dot{\theta} \, U_\infty + 
c^2 \dot{h} \dot{\theta}^2 \theta +
\underline{c^3 \dot{\theta} \ddot{\theta}} +
c^2 \ddot{h} \dot{\theta} +
\underline{c \, \theta \dot{\theta} \, U_\infty^2}
\Big).
\label{eq:power}
\end{aligned}
\end{equation}
In equations \eqref{eq:thrust} and \eqref{eq:power} we underline terms that are inherently out of phase for a linear system with sinusoidal motions, hence we would expect them to be small. However, Liu et al. \cite{liu2014} shows that there may be important nonlinearities in these types of flows and out-of-phase terms are included in Floryan et al. \cite{Floryan2016}; thus we retain them and discuss their importance at the end of the section.

Next we impose sinusoidal motions for heaving and pitching, where pitch has a phase offset $\phi$ from the heave motion, which results in a sinusoidal variation of angle of attack with phase $\psi$:
\begin{align*}
&h = h_0 \sin(2 \pi f t),\quad
\theta = \theta_0 \sin(2 \pi f t+ \phi),\quad
\alpha = \alpha_0 \sin(2 \pi f t+ \psi),
\label{eq:motions}
\end{align*} 
where
\begin{equation}
\alpha_0 = \sqrt{\theta_0^2 + 4 \pi \theta_0 \sin(\phi)\frac{f h_0}{ U_\infty} + \left(2 \pi \frac{f h_0}{U_\infty}\right)^2},
\end{equation}
\begin{equation}
\psi = \arctan \left( \frac{\alpha_0 \sin(\psi)}{\alpha_0 \cos(\psi)} \right) = \arctan \left( \frac{-\theta_0 \sin(\phi) -2 \pi \frac{f h_0}{U_\infty}}{-\theta_0 \cos(\phi)} \right).
\label{eq:psi}
\end{equation} 
To obtain time-averaged quantities, we apply these motion functions to equations \eqref{eq:thrust} and \eqref{eq:power} and integrate with respect to time over one motion cycle. We note the following equalities
\begin{align*}
&\frac{1}{2\pi}\int_0^{2\pi} \sin(t+A)\sin(t+B)dt = \frac{1}{2}\cos(A-B),\\
&\frac{1}{2\pi}\int_0^{2\pi} \sin(t+A)\cos(t+B)dt = \frac{1}{2}\sin(A-B),\\
&\frac{1}{2\pi}\int_0^{2\pi} \cos(t+A)\cos(t+B)dt = \frac{1}{2}\cos(A-B),\\
&\frac{1}{2\pi}\int_0^{2\pi} \cos(t+A)\cos(t+A)\cos(t+B)\sin(t+B) = -\frac{1}{4}\sin(A-B)\cos(A-B),\\
&\frac{1}{2\pi}\int_0^{2\pi} \cos(t+A)\cos(t+B)\cos(t+B)\sin(t+B) = -\frac{1}{8}\sin(A-B),
\end{align*} 
to help us appropriately handle the phase offsets. The resulting time-averaged expressions for thrust and power are
\begin{equation}
\overline{F_x} \sim \rho s c \left(
f \alpha_0 h_0 \sin(\psi) \overline{U_\text{eff}} +
c f^2 \alpha_0 h_0 \cos(\psi) + 
c^2 f^2 \theta_0^2 + 
c f^2 h_0 \theta_0 \cos(\phi) +
c f \theta_0^2 U_\infty
\right)-\overline{F_D},
\label{eq:thrustTA}
\end{equation}
\begin{equation}
\begin{aligned}
\overline{P} \sim &\rho s c\, \Big(   
f \alpha_0 h_0 \sin(\psi) \overline{U_\text{eff}} U_\infty +
c f \alpha_0 \theta_0 \sin(\psi-\phi) \overline{U_\text{eff}}^2 +
c f^2 \alpha_0 h_0 \cos(\psi) U_\infty 
\\
&+ c f^2 \alpha_0 \theta_0 \cos(\psi-\phi)\overline{ U_\text{eff}} +
c^2 f^3 h_0 \theta_0 \sin(\phi) +
c f^3 h_0^2 + 
c f^2 h_0 \theta_0 \cos(\phi) U_\infty
\\
&+ c^3 f^3 \theta_0^2 +
c f \theta_0^2 U_\infty^2 +
c^2 f^3 h_0 \theta_0^3 \sin(\phi) +
c f^3 h_0^2 \theta_0^2 \sin(\phi)\cos(\phi)
\Big).
\label{eq:powerTA}
\end{aligned}
\end{equation}
Henceforth we will neglect the last two terms in the power because they are of higher order and their influence will be small. The exact time-average of the effective velocity of the foil $U_\text{eff}$ is given by the complete elliptic integral of the second kind  $E[\cdot ]$,
\begin{equation*}
\overline{U_\text{eff}} = \frac{2 U_\infty}{\pi} E \left[ -\left( \frac{2 \pi f h_0}{U_\infty} \right)^2 \right].
\end{equation*}
We could use the Taylor series expansion of $E[\cdot ]$ to estimate the above equation, however, for our range of data this is better and more simply approximated by
\begin{equation*}
\overline{U_\text{eff}} \approx U_\infty \sqrt{1+2 \pi^2 \left(\frac{f h_0}{U_\infty} \right)^2},
\end{equation*}

Non-dimensionalizing yields the following thrust and power coefficients
\begin{equation*}
\overline{C_T} \sim
\frac{f \alpha_0 h_0}{U_\infty} U^* \sin(\psi) +
\frac{c f^2 \alpha_0 h_0}{U_\infty^2} \cos(\psi) + 
\frac{c^2 f^2 \theta_0^2}{U_\infty^2} + 
\frac{c f^2 h_0 \theta_0}{U_\infty^2} \cos(\phi) +
\frac{c f \theta_0^2}{U_\infty}
-\overline{C_D},
\end{equation*}
\begin{equation*}
\begin{aligned}
\overline{C_P} &\sim  
\frac{f \alpha_0 h_0}{U_\infty} U^* \sin(\psi) +
\frac{c f \alpha_0 \theta_0}{U_\infty} U^{*2}\sin(\psi-\phi) +
\frac{c f^2 \alpha_0 h_0}{U_\infty^2} \cos(\psi) +
\frac{c^2 f^2 \alpha_0 \theta_0}{U_\infty^2} U^*\cos(\psi-\phi)
\\
&+ \frac{c^2 f^3 h_0 \theta_0}{U_\infty^3} \sin(\phi) +
\frac{c f^3 h_0^2}{U_\infty^3} + 
\frac{c f^2 h_0 \theta_0}{U_\infty^2} \cos(\phi)+
\frac{c^3 f^3 \theta_0^2}{U_\infty^3} +
\frac{c f \theta_0^2}{U_\infty}.
\end{aligned}
\end{equation*}
We will allow the drag term $\overline{C_D}$ to be a linear function of pitch amplitude $\theta_0$, that is, a linear function of the projected frontal area for slow motions ($f \rightarrow 0$). As shown in Appendix B, this is a fair approximation. We can express the expressions for the thrust and power coefficients in terms of the non-dimensional parameters for each independent motion, i.e., the Strouhal numbers $St_h = 2f h_0/U_\infty$ and $St_\theta = 2 f c\theta_0/U_\infty$, reduced frequency $f^*=fc/U_\infty$, and the amplitude to chord ratios $a_h^*=h_0/c$ and $a_\theta^*=\theta_0$.  The expressions for thrust and power become
\begin{equation}
\overline{C_T} \sim
\underbrace{\alpha_0 St_h U^* \sin(\psi) \rule[-12pt]{0pt}{5pt}}_{\mbox{1}} +
\underbrace{\alpha_0 f^* St_h \cos(\psi) \rule[-12pt]{0pt}{5pt}}_{\mbox{2}} + 
\underbrace{St_\theta^2 \rule[-12pt]{0pt}{5pt}}_{\mbox{3}} + 
\underbrace{St_h St_\theta \cos(\phi) \rule[-12pt]{0pt}{5pt}}_{\mbox{4}} +
\underbrace{St_\theta a_\theta^* \rule[-12pt]{0pt}{5pt}}_{\mbox{5}}
-\underbrace{a_\theta^* \rule[-12pt]{0pt}{5pt}}_{\mbox{6}},
\label{eq:thrustCoeff}
\end{equation}
\begin{equation}
\begin{aligned}
\overline{C_P} &\sim    
\alpha_0 St_h U^* \sin(\psi) +
\alpha_0 St_\theta U^{*2}\sin(\psi-\phi) +
\alpha_0 f^* St_h \cos(\psi) +
\alpha_0 f^* St_\theta U^*\cos(\psi-\phi)
\\
&+ f^* St_h\, St_\theta \sin(\phi) +
f^* St_h^2 + 
St_h\, St_\theta \cos(\phi) + 
f^* St_\theta^2 +
St_\theta\, a_\theta^*.
\label{eq:powerCoeff}
\end{aligned}
\end{equation}
We can attribute the following physical mechanisms to thrust production in equation \eqref{eq:thrustCoeff}: steady and unsteady aerodynamic lift (terms 1 and 2, respectively); partly foil acceleration and centrifugal force (term 3, which is a combination of multiple earlier terms); partly foil acceleration and Coriolis force (term 4, which is a combination of multiple earlier terms); solely Coriolis force (term 5); and viscous drag (term 6). As a check on our work, in the special cases of pure heave ($\theta_0=0$) and pure pitch ($h_0=0$) we arrive at the expressions provided by Floryan et al. \cite{Floryan2016}. 

Further simplification is possible, though we will make simplifications that go beyond those of Floryan et al. \cite{Floryan2016}. First, we apply equation \eqref{eq:psi} to equations \eqref{eq:thrustCoeff} and \eqref{eq:powerCoeff} by recognizing that $\alpha_0 \cos(\psi) = -\theta_0 \cos(\phi)$ and $\alpha_0 \sin(\psi) = -\theta_0 \sin(\phi) - 2\pi\,f\,h_0/U_\infty$. Second, when it is possible we will combine terms into the total motion Strouhal number $St=2f a_0/U_\infty$ where $a=h+c\theta$, thus $St^2=St_h^2+St_\theta^2+2St_h St_\theta cos(\phi)$. Third, we will recognize that $U^* \approx 1$ is a fair approximation, as changes in $U^*$ are relatively small when compared to other input parameters like $St$, $f^*$, or $a^*$. Although this approximation may seem aggressive, our results indicate that inclusion of $U^*$ has a minor impact on the scaling. With these changes, we arrive at the final expressions for the thrust and power coefficients:
\begin{equation}
\overline{C_T} =
c_1\, St^2 +
c_2\, St_h a_\theta^* \sin(\phi) +
\underline{c_3\, St_\theta a_\theta^*}
- c_4\, a_\theta^* ,
\label{eq:finalCT}
\end{equation}
\begin{equation}
\begin{aligned}
\overline{C_P} &=    
c_5\, St^2 +
c_6\, f^* St_h St_\theta \sin(\phi) +
c_7\, St_h a_\theta^* \sin(\phi) +
\underline{c_8\, f^* St_h^2} + 
\underline{c_9\, f^* St_\theta^2} +
\underline{c_{10}\, St_\theta\, a_\theta^*},
\label{eq:finalCP}
\end{aligned}
\end{equation}
where we we have applied coefficients $c_n$ to each term to account for the constants that have been dropped throughout our analysis. These coefficients are determined via linear regression over the entire experimental dataset. Note that the value of the coefficients may differ for other situations, such as a change in the airfoil type, and can be more specifically tuned for a refined dataset, like a fixed pitch phase angle $\phi$. The purpose of this analysis is to show the physical phenomena that govern the thrust and power, which is expressed by the sources of the terms in the scaling relations.

\begin{figure}
\centering
\includegraphics[width=1\textwidth]{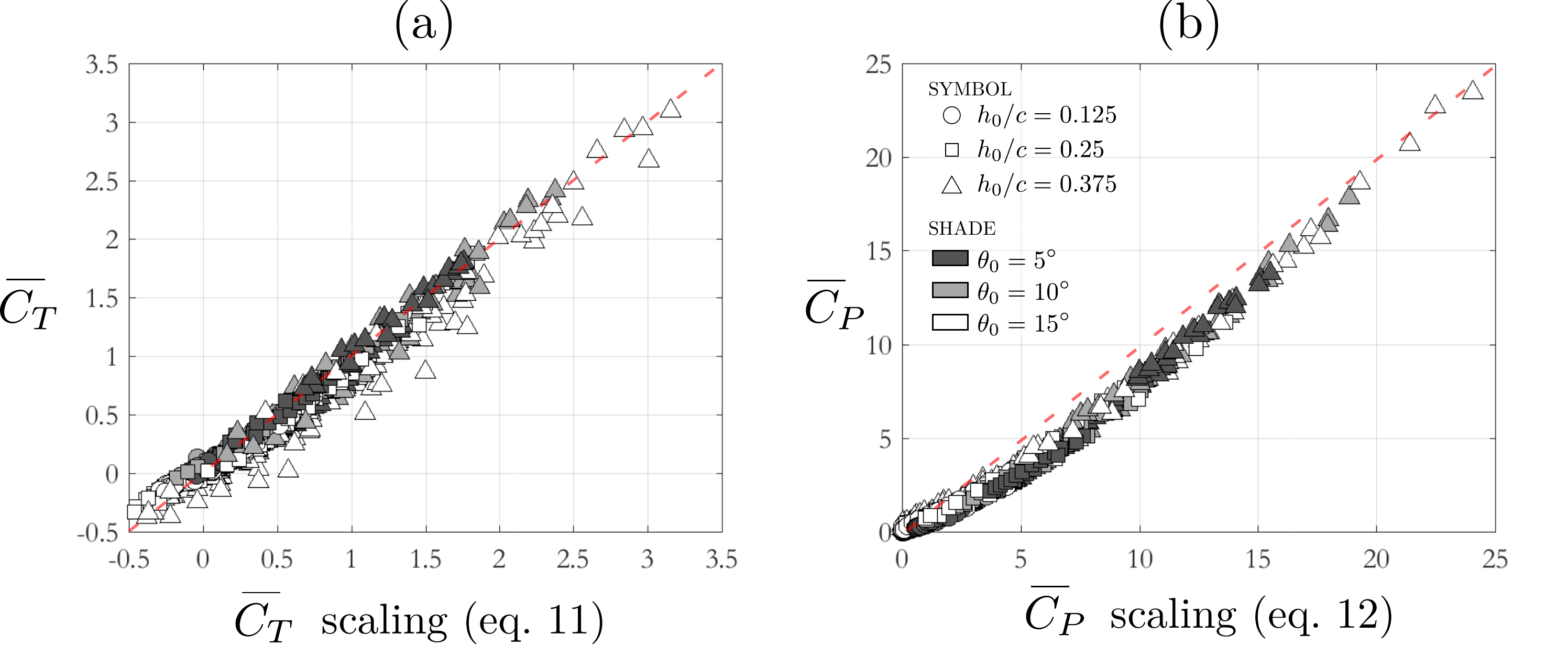}
\caption{Scaling of the time-averaged (a) thrust and (b) power coefficients for all motion amplitudes and phases tested (see table \ref{tab:Cases}). Thrust constants: $c_1=4.84$, $c_2=-5.96$, $c_3=-2.82$, $c_4=0.48$; power constants: $c_5=25.1$, $c_6=32.1$, $c_7=4.95$, $c_8=41.35$, $c_9=14.98$, $c_{10}=-25.77$.}
\label{fig:scalingP}
\end{figure}

The experimental data for all motion amplitudes and phases, plotted as a function of these scaling relations, are shown in figure \ref{fig:scalingP}. For both thrust and power, we see collapse of the data, indicating that our simplified model closely describes the propulsive performance for these types of foil motions. A linear collapse of the data is ideal, as we see in the thrust, but the power shows slight nonlinear behavior (also seen in \cite{Floryan2016}). We could completely linearize the power scaling by adding a single higher order term $St^3$, however, we do not find this term in our analysis and cannot physically describe its existence, thus we do not include it.

In equations \eqref{eq:finalCT} and \eqref{eq:finalCP} we have underlined the terms that come from inherently out-of-phase motions, and their necessity may reveal flow physics. The phase differences that cause out-of-phase terms to be important could be due to the influence of the wake on the foil. Scherer \cite{scherer1968} explains that the induced velocity of the wake on the foil changes the instantaneous angle of attack and inertial forces from added mass, which is equivalent to a time-lag on the circulatory forces in the linearized theory (for a nonlinear system this equivalency cannot be made \cite{liu2014}). We find the heave-based term in the power $f^*St_h^2$, which was originally sourced from a $\ddot{h}\dot{h}$ term in equation \eqref{eq:power}, to be critical in the scaling. However, the other pitch-based terms in both the thrust and power could be neglected without much penalty on the data collapse. This could be because the out-of-phase pitch terms are all of the order $\mathcal{O}(\theta_0^2)$, so they will be smaller in their relative contributions.

\begin{figure}
\centering
\includegraphics[width=1\textwidth]{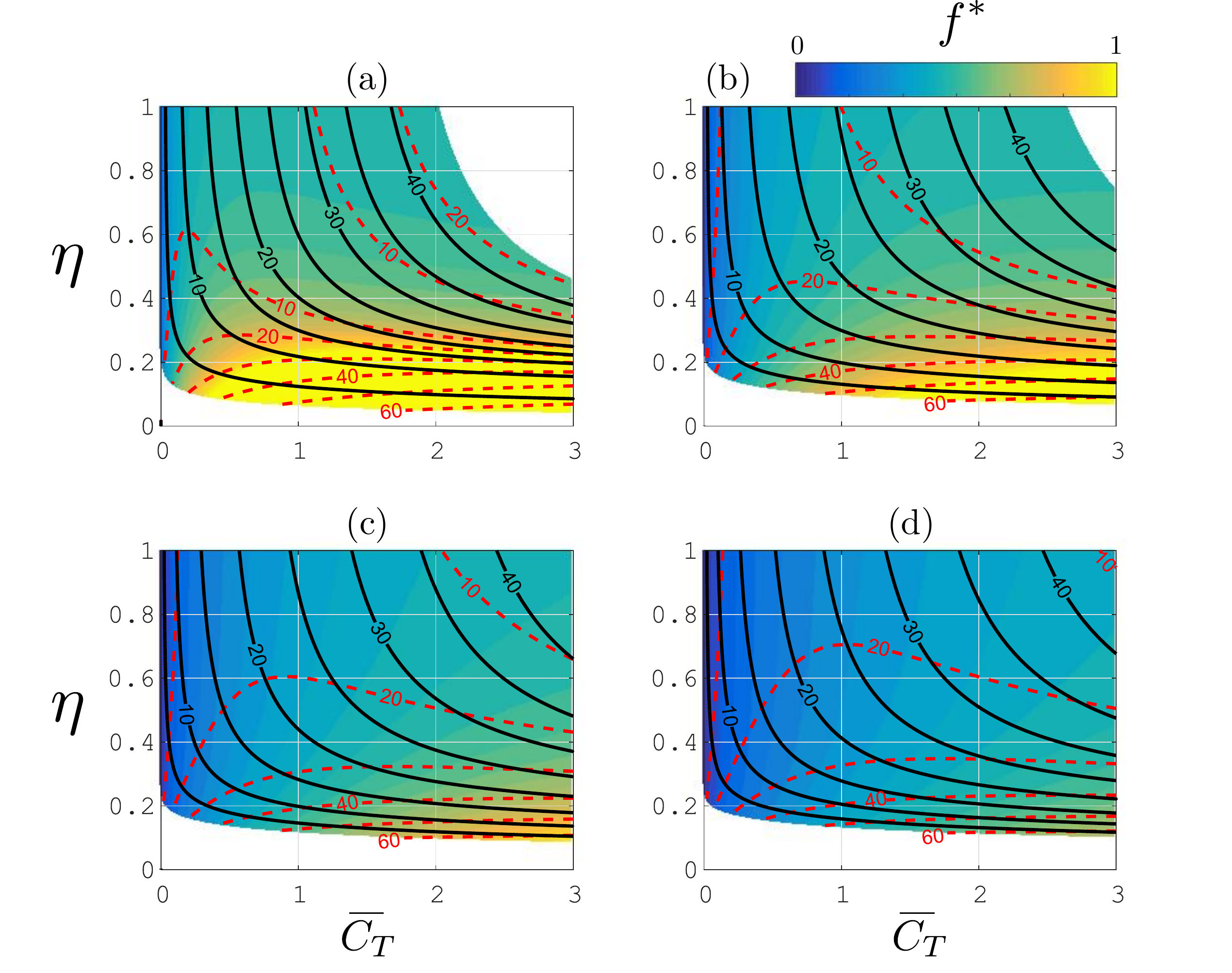}
\caption{Performance maps of efficiency and time-averaged thrust coefficient for a heaving and pitching foil with phase offset $\phi=270^\circ$. Contours of pitch amplitude (black solid lines) and peak angle of attack (red dashed lines) superimposed on a color contour of reduced frequency. Heave amplitude to chord ratio (a) $h_0/c=0.125$, (b) 0.25, (c) 0.375, (d) 0.5. Calculated from equations \eqref{eq:finalCT} and \eqref{eq:finalCP}, with constants from figure \ref{fig:scalingP}.}
\label{fig:perfSpace}
\end{figure}

Figure \ref{fig:perfSpace} shows performance maps of efficiency versus mean thrust with contours of pitch angle, peak angle of attack, and reduced frequency for multiple heave amplitudes, calculated from equations \eqref{eq:finalCT} and \eqref{eq:finalCP}. When creating these performance maps, we have neglected the drag term in the scaling relation for thrust in order to calculate ``ideal'' efficiencies, that is, we have removed the effect where efficiency rapidly decays at low reduced frequency (see figure~\ref{fig:pitchHeave} and the accompanying discussion). The maps extend a good deal beyond our experimental data set, and although we cannot validate these extrapolated regions (especially where our approximations break down), they may be useful in identifying trends and regions which should be explored further. Propulsors that produce high thrust and do so efficiently appear in the top-right corner of the performance maps, so this is where propulsors should operate. There are two standout trends: (1) for fixed motion amplitudes, decreasing the peak angle of attack increases efficiency (similar behavior seen in \cite{kaya2007}); and (2) for fixed frequency and heave amplitude, increasing the pitch amplitude increases both efficiency and thrust. In other words, bigger, slower motions improve efficiency. 

A discussion in \cite{alexander2003} illuminates why bigger, slower motions improve efficiency. For unsteady propulsion, the thrust is equal to the rate at which streamwise momentum is added to the wake, $u_w \dot{m}_w$, where $m_w$ is the mass of water that is accelerated from rest to velocity $u_w$. The same thrust can be generated by accelerating a small mass of water to a high velocity (small, fast motions), or a large mass of water to a low velocity (big, slow motions). 
The power that the system has to expend in order to overcome body drag during steady swimming is the product of drag and speed: $D U_\infty = u_w \dot{m}_w U_\infty$. We think of this as the useful power. The total power expended by the system includes the useful power and the power lost to the wake, equal to the rate at which the kinetic energy of the wake is increased: $\frac{1}{2} u_w^2 \dot{m}_w$. The efficiency of the system is the ratio of the useful to total power, and is
\begin{equation*}
\eta = \frac{U_\infty}{U_\infty +\frac{1}{2} u_w}.
\end{equation*}
This simple analysis suggests that for more efficient motions, it is desirable to minimize $u_w$. Thus, big slow motions should be more efficient than small fast motions that produce equal thrust.

\subsection{Conclusions}

A heaving and pitching teardrop foil was studied experimentally in an effort to maximize propulsive performance, that is, thrust and efficiency. Combining heave and pitch motions generally achieves improved performance compared to heave or pitch in isolation. A critical parameter is the phase difference between the heave and pitch motions. Peak thrust occurred near $\phi=330^\circ$, minimum power near $\phi=210^\circ$, and peak efficiency near $\phi=270^\circ$, which was coincident with maintaining the smallest peak angles of attack.  At $\phi=270^\circ$, we found that the component of the power required to turn the foil is actually negative in the mean, indicating that at this phase difference the fluid is doing work to help the motion.

Scaling relations for the mean thrust and power generated by a heaving and pitching foil were developed by considering lift-based and added mass forces and moments, based on the relations for heaving \emph{or} pitching foils developed by Floryan et al. \cite{Floryan2016}. These relations describe the experimental performance behavior well and provide a valuable guide for further improving performance. Generally, the scaling relations indicate that, to increase thrust and efficiency, we need to increase the motion amplitudes (specifically pitch angle) while minimizing the peak angle of attack. 

The results of \cite{scherer1968, anderson1998, read2003} on large-amplitude heaving and pitching foils lend confidence to our conclusions; they used relatively large motions ($h_0/c$ up to 0.75 and $\theta_0$ up to $65^\circ$) compared to the experiments described here, while keeping the angle of attack small ($\alpha_0 \approx 20^\circ$), leading to very high propulsive efficiencies, $\eta \approx 0.55-0.8$. For the present experiments on a teardrop airfoil, efficiencies of 45--50\% were obtained over a large parameter space of motion amplitudes. Some cases displayed efficiencies as high as 60--75\%, but the peak values are very sensitive to the drag on the foil, and so the foil profile is likely to be a crucial design parameter.  Such considerations will be left for future study.

\section*{Acknowledgments}
This work was supported by ONR grant N00014-14-1-0533 (Program Manager R. Brizzolara). 

\section*{References}

\bibliographystyle{aiaa}
\bibliography{BibliographyLibrary}

\section*{Appendix}\label{S5S1}
\setcounter{subsection}{0}
\subsection{Complete performance results}
\begin{figure}
\includegraphics[width=0.8\textwidth]{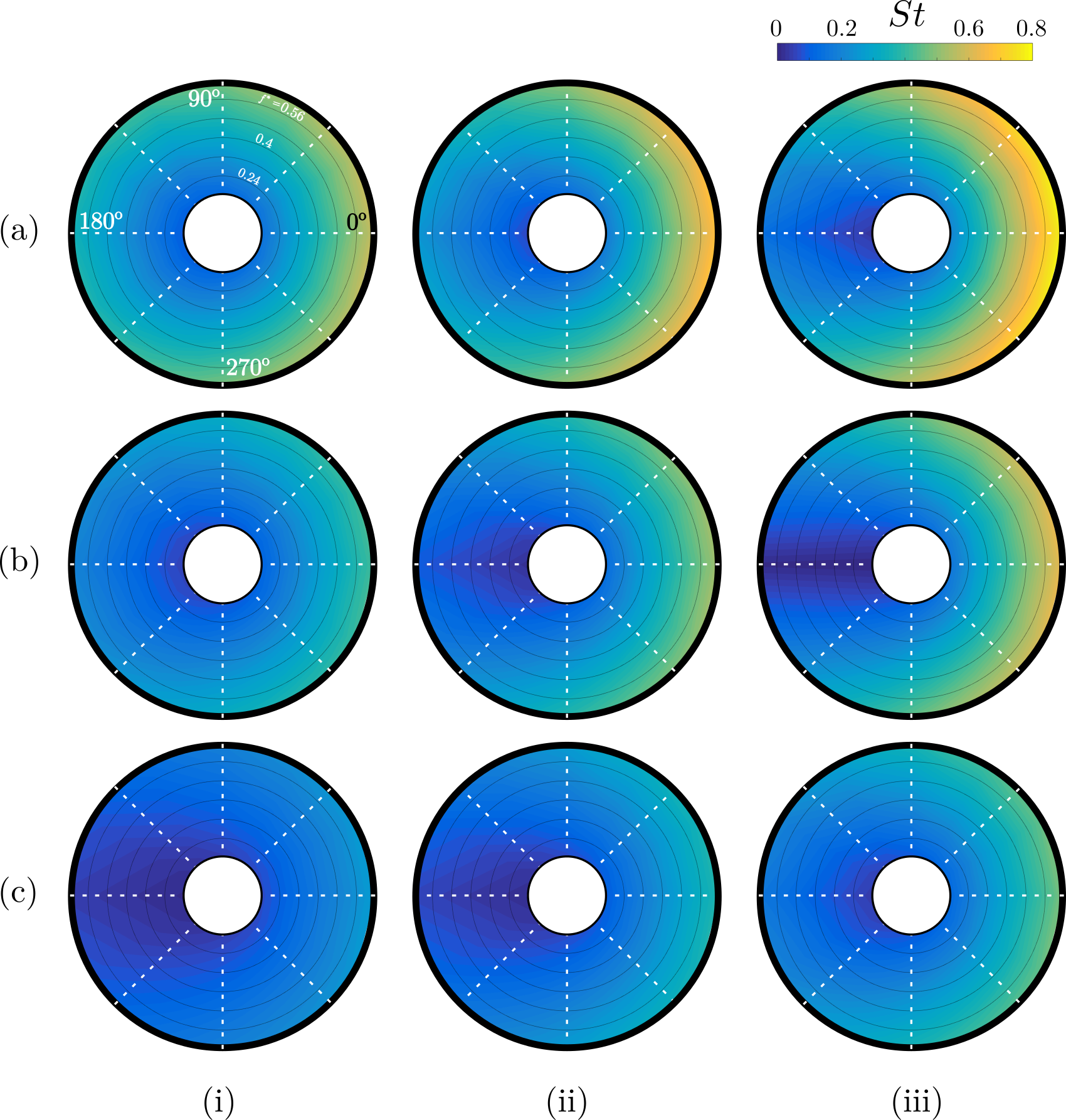}
\caption{Impact of phase offset ($\phi$, azimuthal axis) and reduced frequency ($f^*$, radial axis) on Strouhal number for the following combinations of heave amplitudes (a) $h_0/c=0.375$ (b) $0.25$, (c) $0.125$, and pitch amplitudes (i) $\theta_0=5^\circ$, (ii) $10^\circ$, (iii) $15^\circ$. Frequency increases radially outward with lines marking levels at $f^*=0.16, 0.24,...0.64$.}
\label{fig:phase1}
\end{figure}

\begin{figure}
\includegraphics[width=0.8\textwidth]{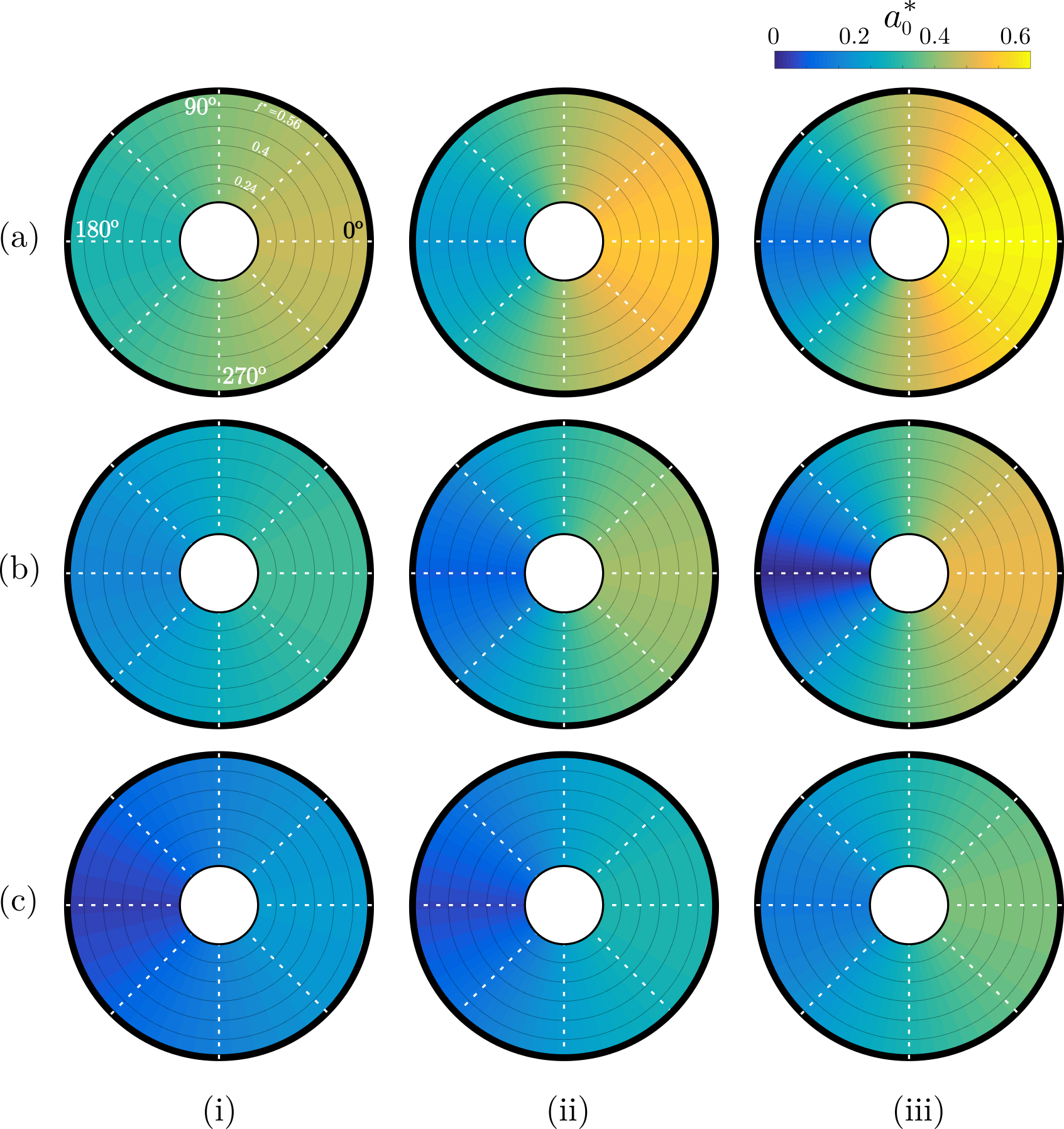}
\caption{Impact of phase offset ($\phi$, azimuthal axis) and reduced frequency ($f^*$, radial axis) on the peak trailing edge amplitude to chord ratio for the following combinations of heave amplitudes (a) $h_0/c=0.375$ (b) $0.25$, (c) $0.125$, and pitch amplitudes (i) $\theta_0=5^\circ$, (ii) $10^\circ$, (iii) $15^\circ$. Frequency increases radially outward with lines marking levels at $f^*=0.16, 0.24,\ldots, 0.64$.}
\label{fig:phase2}
\end{figure}

\begin{figure}
\includegraphics[width=0.8\textwidth]{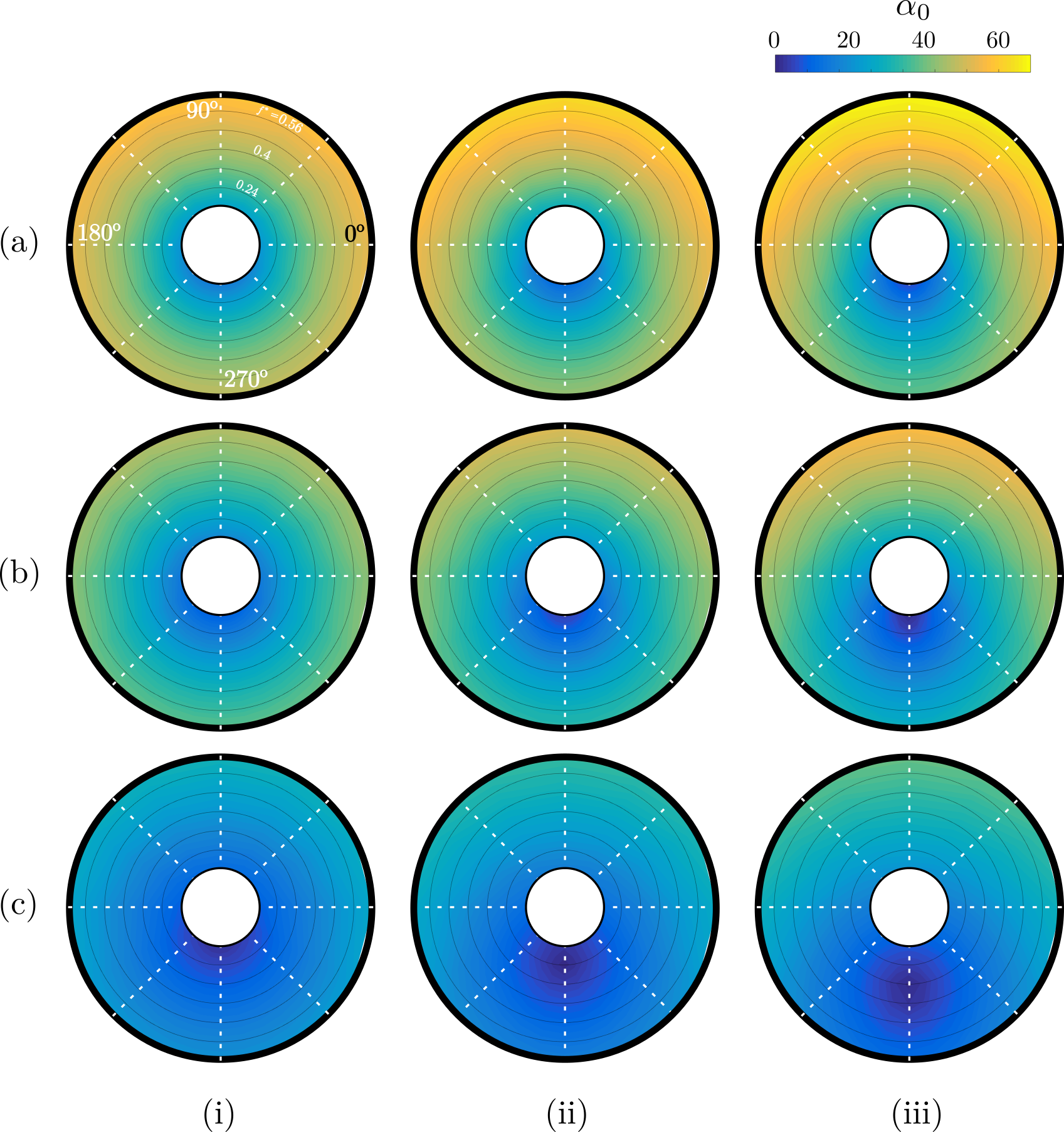}
\caption{Impact of phase offset ($\phi$, azimuthal axis) and reduced frequency ($f^*$, radial axis) on the peak angle of attack for the following combinations of heave amplitudes (a) $h_0/c=0.375$ (b) $0.25$, (c) $0.125$, and pitch amplitudes (i) $\theta_0=5^\circ$, (ii) $10^\circ$, (iii) $15^\circ$. Frequency increases radially outward with lines marking levels at $f^*=0.16, 0.24,\ldots, 0.64$.}
\label{fig:phase3}
\end{figure}

\begin{figure}
\includegraphics[width=0.8\textwidth]{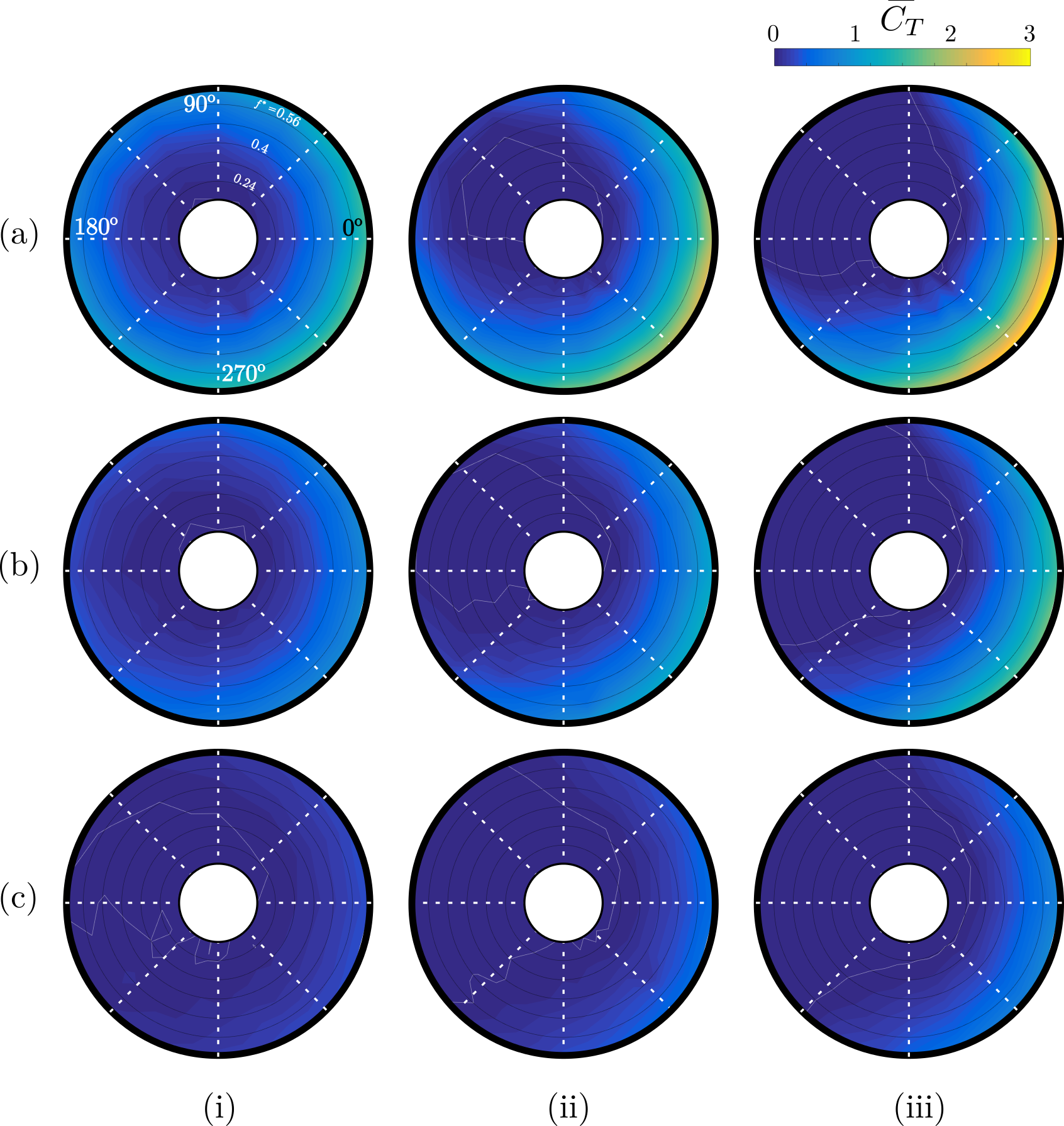}
\caption{Impact of phase offset ($\phi$, azimuthal axis) and reduced frequency ($f^*$, radial axis) on the thrust coefficient for the following combinations of heave amplitudes (a) $h_0/c=0.375$ (b) $0.25$, (c) $0.125$, and pitch amplitudes (i) $\theta_0=5^\circ$, (ii) $10^\circ$, (iii) $15^\circ$. Frequency increases radially outward with lines marking levels at $f^*=0.16, 0.24,\ldots, 0.64$.}
\label{fig:phase4}
\end{figure}

\begin{figure}
\includegraphics[width=0.8\textwidth]{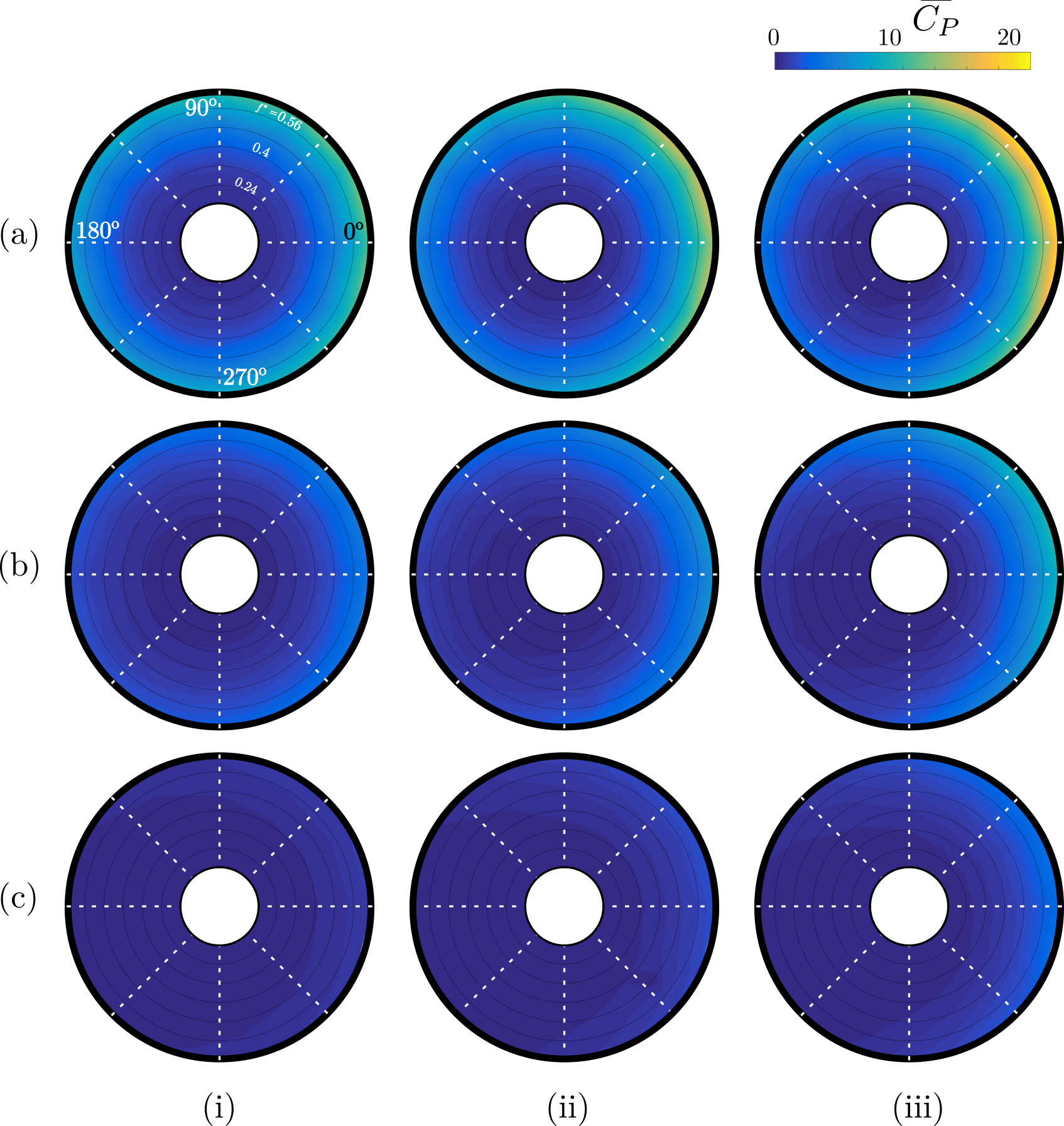}
\caption{Impact of phase offset ($\phi$, azimuthal axis) and reduced frequency ($f^*$, radial axis) on the power coefficient for the following combinations of heave amplitudes (a) $h_0/c=0.375$ (b) $0.25$, (c) $0.125$, and pitch amplitudes (i) $\theta_0=5^\circ$, (ii) $10^\circ$, (iii) $15^\circ$. Frequency increases radially outward with lines marking levels at $f^*=0.16, 0.24,\ldots, 0.64$.}
\label{fig:phase5}
\end{figure}

\begin{figure}
\includegraphics[width=0.8\textwidth]{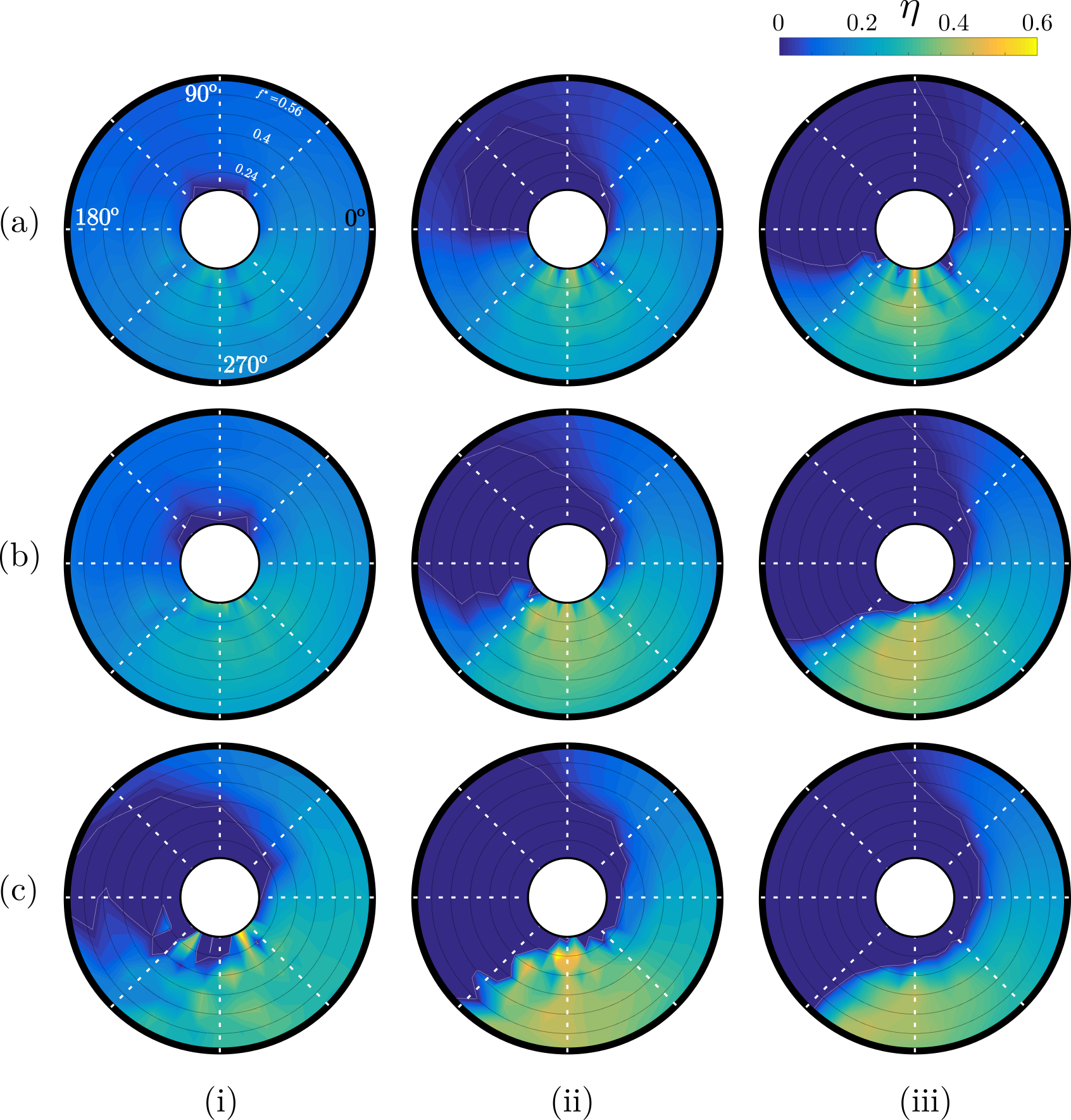}
\caption{Impact of phase offset ($\phi$, azimuthal axis) and reduced frequency ($f^*$, radial axis) on the propulsive efficiency for the following combinations of heave amplitudes (a) $h_0/c=0.375$ (b) $0.25$, (c) $0.125$, and pitch amplitudes (i) $\theta_0=5^\circ$, (ii) $10^\circ$, (iii) $15^\circ$. Frequency increases radially outward with lines marking levels at $f^*=0.16, 0.24,\ldots, 0.64$.}
\label{fig:phase6}
\end{figure}

Figures \ref{fig:phase1}--\ref{fig:phase6} show the non-dimensional input parameters (Strouhal number, trailing edge amplitude to chord ratio, and peak angle of attack) and time-averaged output performance (thrust, power, efficiency) for all the phase offsets, frequencies, heave amplitudes, and pitch amplitudes tested (see table \ref{tab:Cases}).

\subsection{Foil viscous drag assessment}
\begin{figure}
\includegraphics[width=0.8\textwidth]{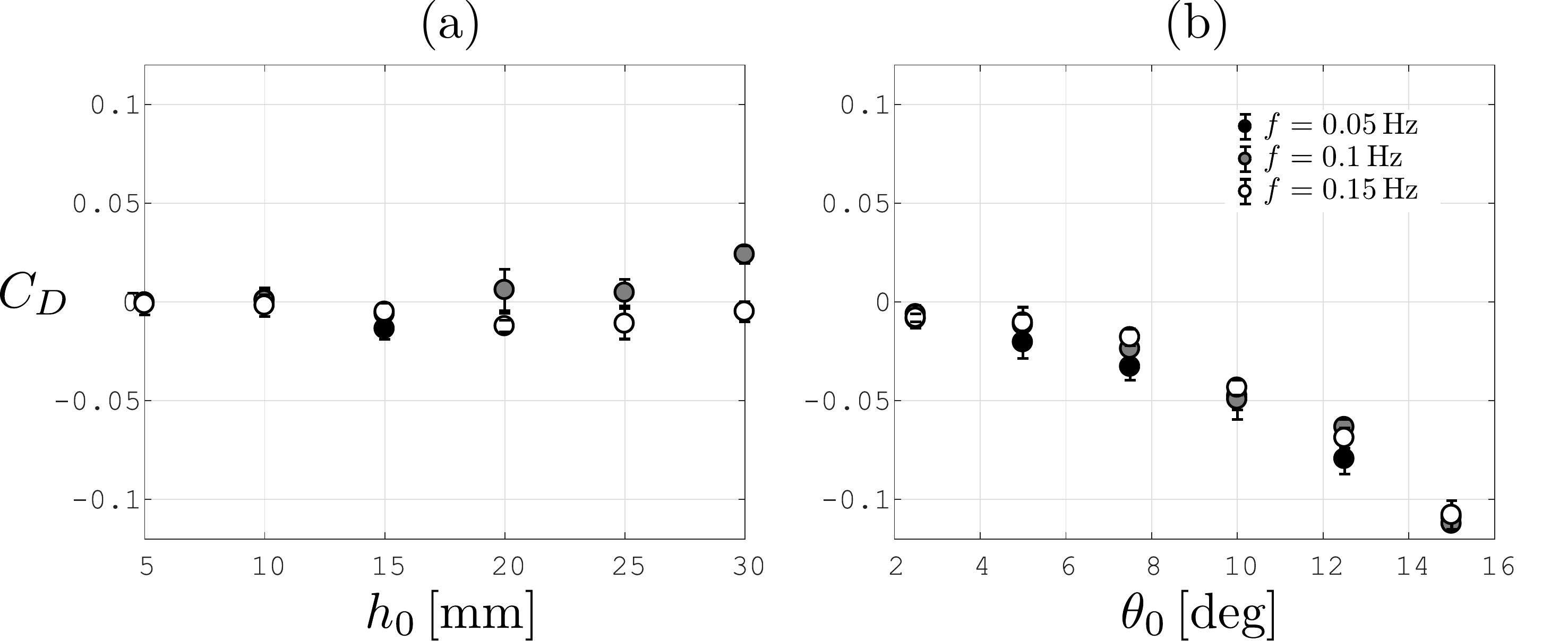}
\caption{Drag coefficient as it varies with frequency and motion amplitude for (a) heaving and (b) pitching motions.}
\label{fig:cd}
\end{figure}

As the foil becomes still ($f\rightarrow 0$), the viscous drag of the foil should depend on the mean projected area of the foil over the entire motion cycle. Accordingly, the drag should only be a function of the pitch amplitude and not a function of the heave amplitude. Figure \ref{fig:cd} shows the drag coefficient as it varies with motion amplitude for heaving and pitching motions at three very low frequencies. The forces due to viscous drag are low, thus the experiments were repeated 10 times each and the uncertainty bars on the symbols reflect the standard deviation of the data. Note that points at heave amplitudes $h_0=$ 20, 25, and 30 mm for frequency $f=$ 0.05 Hz were removed because the linear actuator could not operate smoothly at this combination of low frequency/high amplitude. The drag coefficient is clearly a function of the pitch but not the heave amplitude for steady motion.

\end{document}